\documentclass[a4paper,superscriptaddress,twocolumn,pra]{revtex4-1}
\usepackage{graphicx}
\usepackage{latexsym}
\usepackage{amsmath}
\usepackage{amssymb}

\begin{document}

\title{Quantum discord in nuclear magnetic resonance systems at room temperature}

\author{J. Maziero}
\email{jonasmaziero@gmail.com}
\affiliation{Universidade Federal do Pampa, Campus Bag\'{e}, 96413-170 Bag\'{e}, Rio Grande do Sul, Brazil}

\author{R. Auccaise}
\affiliation{Empresa Brasileira de Pesquisa Agropecu\'{a}ria, Rua Jardim Bot\^{a}nico 1024, 22460-000 Rio de Janeiro, Rio de Janeiro, Brazil}

\author{L. C. C\'eleri}
\affiliation{Instituto de F\'{i}sica, Universidade Federal de Goi\'{a}s, 74.001-970 Goi\^{a}nia, Goi\'{a}s, Brazil}

\author{D. O. Soares-Pinto}
\affiliation{Instituto de F\'{\i}sica de S\~{a}o Carlos, Universidade de S\~{a}o Paulo, P.O. Box 369, 13560-970 S\~{a}o Carlos, S\~{a}o Paulo, Brazil}

\author{E. R. deAzevedo}
\affiliation{Instituto de F\'{\i}sica de S\~{a}o Carlos, Universidade de S\~{a}o Paulo, P.O. Box 369, 13560-970 S\~{a}o Carlos, S\~{a}o Paulo, Brazil}

\author{T. J. Bonagamba}
\affiliation{Instituto de F\'{\i}sica de S\~{a}o Carlos, Universidade de S\~{a}o Paulo, P.O. Box 369, 13560-970 S\~{a}o Carlos, S\~{a}o Paulo, Brazil}

\author{R. S. Sarthour}
\affiliation{Centro Brasileiro de Pesquisas F\'{\i}sicas, Rua Dr. Xavier Sigaud 150, 22290-180 Rio de Janeiro, Rio de Janeiro, Brazil}

\author{I. S. Oliveira}
\affiliation{Centro Brasileiro de Pesquisas F\'{\i}sicas, Rua Dr. Xavier Sigaud 150, 22290-180 Rio de Janeiro, Rio de Janeiro, Brazil}

\author{R. M. Serra}
\affiliation{Centro de Ci\^encias Naturais e Humanas, Universidade Federal do ABC, R. Santa Ad\'elia 166, 09210-170 Santo Andr\'e, S\~ao Paulo, Brazil}

\begin{abstract}
We review the theoretical and the experimental aspects regarding the quantification and identification of quantum correlations in liquid-state nuclear magnetic resonance (NMR) systems at room temperature. We start by introducing a formal method to obtain the quantum discord and its classical counterpart in systems described by a deviation matrix. Next, we apply such a method to experimentally demonstrate that the peculiar dynamics, with a sudden change behaviour, of quantum discord under decoherence, theoretically predicted only for phase-noise channels, is also present even under the effect of a thermal environment. This result shows that such a phenomena are much stronger than we could think, at principle. Walking through a different path, we discuss an observable witness for the quantumness of correlations in two-qubit systems and present the first experimental implementation of such a quantity in a NMR setup. Such a witness could be very useful in situations were the knowledge of the nature of correlations (in contrast of how much correlations) presented in a given state is enough.
\end{abstract}

\maketitle

\section{Introduction}

\subsection{Bell's Inequalities, Entanglement and the No-Local-Broadcasting Theorem}

The importance of the search for a deep understanding about the quantum nature of correlations among the parts constituents of a composite system began to be recognized with the seminal works published by Einstein, Podolsky and Rosen \cite{EPR} and Schr\"odinger \cite{Schrodinger} in 1935. Recently, developments in the study of quantum correlations have been linked with advances in quantum information science. Initially, the only remarkable aspect of quantum correlations was their non-local nature, named entanglement by Schr\"odinger \cite{Schrodinger} and mathematically formalized by Bell in 1964 \cite{Bell64}. Bell proved that certain relations among the average values of observables of a composite system must be satisfied under the assumptions of realism and locality. However, such inequalities could be violated for quantum systems in certain entangled states. These relations are known, in general, as Bell's inequalities. States violating a Bell's inequality are said to possess non-local quantum correlations \cite{Clauser-Shimony,Aspect,Mermin}.

In 1989, Werner \cite{Werner89} reported a groundbreaking result adding a new perspective to the study of non-classicality of the correlations in multipartite systems. By defining entanglement operationally as those correlations that cannot be generated through local quantum operations assisted by classical communication, Werner noticed the existence of entangled states that do not violate any Bell's inequality. Surprisingly, Popescu showed that some non-separable-local states can be used to do quantum teleportation with fidelity greater than what is possible by classical means \cite{Popescu}. 

Recall that, in quantum mechanics, the more general quantum operation describing changes between quantum states can be written, in the operator-sum representation, as \cite{Nielsen-Chuang}
\begin{equation*}
\mathcal{E}(\rho)=\sum_{j}K_{j}\rho K_{j}^{\dagger},
\end{equation*}
where $K_{j}$ are the Kraus' operators acting on the state space of the system, $\mathcal{H}$, satisfying the relation 
\begin{equation*}
\sum_{j}K_{j}^{\dagger}K_{j}\le\mathbb{I},
\end{equation*} 
with $\mathbb{I}$ being the identity operator in $\mathcal{H}$. The equality in this relation is a sufficient condition for the map $\mathcal{E}:\mathcal{H} \rightarrow\mathcal{H}$ to be a trace-preserving one, that is, $\mathrm{Tr}\left[\mathcal{E}(\rho)\right] = 1$. 

It turns out that the most general bipartite state that can be created via local quantum operations ($\mathcal{E}_{A}$ and $\mathcal{E}_{B}$) and classical communication is of the form
\begin{equation}
\rho_{AB}^{sep}=\sum_{j}p_{j}\rho_{j}^{A}\otimes\rho_{j}^{B},
\label{eq:sep-rho}
\end{equation}
where $\{p_{j}\}$ is a probability distribution (i.e. $p_{j}\ge0$ $\sum_{j}p_{j}=1$) and $\rho_{j}^{A(B)}\in\mathcal{H}_{A(B)}$ is the reduced density operator of subsystem $A\mbox{ }(B)$. States that can be written in the form shown in Eq. (\ref{eq:sep-rho}) are said to be separable. All the states that are not separable are defined as being entangled (or non-separable).

The non-local and non-separable aspects of the correlations in a composite system, as discussed above, are distinguishing features of the quantum realm. Nevertheless, we believe that, in a more general fashion, it is the possibility of locally redistribute the correlations in a multipartite state that broadly defines its nature. In 2008, Piani and co-authors proved the so called no-local-broadcasting theorem \cite{Piani-Horodecki}, showing that the correlations in a bipartite state $\rho_{AB}$ can be redistributed via local quantum operations if and only if this state can be cast in the following form:
\begin{equation}
\rho_{AB}^{cc}=\sum_{j,k}p_{jk}|\psi_{j}^{A}\rangle\langle\psi_{j}^{A}|\otimes|\phi_{k}^{B}\rangle\langle\phi_{k}^{B}|,
\label{eq:CC-states}
\end{equation}
where $\{p_{jk}\}$ is a probability distribution and $\{|\psi_{j}^{A}\rangle\}$ ($\{|\phi_{k}^{B}\rangle\}$) is an orthonormal basis for subsystem $A$ ($B$). By comparing this equation with Eq. (\ref{eq:sep-rho}) we see that $\rho_{AB}^{cc}$ is a subset of the separable states (which forms a convex set). Therefore, even separable states can present quantum correlations, in the sense that such correlations cannot be locally broadcast. This kind of non-classical correlation can be indicated, and in some sense quantified, by a quantity called quantum discord (QD) \cite{Celeri-Review,Modi-Review}, that will be discussed in more detail in the next subsection.

\subsection{Quantum Discord}

In this article we will focus on QD. This choice is motivated by the fact that this measure was the first one proposed in literature and was studied in most of the works presented so far. A detailed discussion about other measures of quantum correlations, as well as several other aspects of such measures, is presented in the reviews \cite{Celeri-Review,Modi-Review}. 

Before discussing QD it is instructive to briefly review some aspects of classical information theory, in order to set the notation and define the relevant quantities.

\subsubsection{Some Concepts from Classical Information Theory}

In classical information theory, the uncertainty about a random variable $A$ is quantified by its Shannon's entropy \cite{Shannon}
\begin{equation}
H(A):=-\sum_{a}p_{a}\log_{2}p_{a},
\label{eq:Shannon-H}
\end{equation}
with $p_{a}:=\mathrm{Pr}(A=a)$ standing for the probability that $A$ assuming the value $a$. The binary logarithm is used here such that information is measured in bits.

Two random variables are said to be correlated if they share some information. Thus, by knowing the value of one of them, say $B$, we can acquire information about the other, $A$. The difference in the uncertainty about $A$ before and after we know $B$,
\begin{equation}
J(A\mbox{:}B):=H(A) - H(A|B),
\label{eq:CMI-J}
\end{equation}
is a measure of the correlations between the two random variables ($A$ and $B$). This quantity is called mutual information. In the above equation the conditional entropy reads
\begin{equation*}
 H(A|B):=-\sum_{a,b}p_{a,b}\log_{2}p_{a|b},
\end{equation*}
where $p_{a,b}:=\mathrm{Pr}(A=a,B=b)$ and, from the definition for the conditional probability for $A$ to assume the value $a$ when $B$ is equal to $b$, $p_{a|b}:=p_{a,b}/p_{b}$, it follows that
\begin{equation*}
 H(A|B)=H(A\text{,}B)-H(B),  
\end{equation*}
with $H(A,B)=-\sum_{a,b}p_{a,b}\log_{2}p_{a,b}$. Hence an equivalent expression for the mutual information between $A$ and $B$ can be written as
\begin{equation}
I(A\mbox{:}B)=H(A)+H(B)-H(A\text{,}B).
\label{eq:CMI-I}
\end{equation}
In classical information theory, the relation
\begin{equation*}
J(A\mbox{:}B)=I(A\mbox{:}B)  
\end{equation*}
is always valid. However, the extensions to the quantum realm of these two, classically equivalent, expressions for the mutual information differ in general. Such non-equivalence is at the heart of the quantum discord, that turns out to be a measure of quantum aspects of correlation, as will be discussed next.

\subsubsection{The Original Definition of Quantum Discord}

The uncertainty about a system $S$, described by a density operator $\rho_{S}$, is quantified in quantum information theory through its von Neumann's entropy \cite{Schumacher}:
\begin{equation}
S(\rho_{S}):=-\mathrm{Tr}(\rho_{S}\log_{2}\rho_{S}).
\label{eq:vonNeumann-S}
\end{equation}
As a result, a direct extension for the mutual information shown in Eq. (\ref{eq:CMI-I}) can be obtained as
\begin{equation}
\mathcal{I}(\rho_{AB}):=S(\rho_{A})+S(\rho_{B})-S(\rho_{AB}).
\label{eq:QMI-I}
\end{equation}
This quantity is named quantum mutual information and it is regarded as a measure of the total correlations between subsystems $A$ and $B$ when the state of the joint system is $\rho_{AB}$ \cite{Groisman,Schumacher-I}.

With the purpose of obtaining a quantum extension for the mutual information in Eq. (\ref{eq:CMI-J}), let us consider the measurement of an observable represented by the following Hermitian operator
\begin{equation*}
O_{B}=\sum_{j}o_{j}|\Pi_{j}^{B}\rangle\langle\Pi_{j}^{B}|, 
\end{equation*}
that is defined on the state space of subsystem $B$, $\mathcal{H}_{B}$. If the system is initially in state $\rho_{AB}$, the value $o_{j}$ is obtained with probability
\begin{equation*}
 p_{j}=\mathrm{Tr}(\mathbb{I}_{A}\otimes\Pi_{j}^{B}\rho_{AB}) 
\end{equation*}
and the state of subsystem $A$, immediately after the measurement, reads
\begin{equation*}
 \rho_{j}^{A}=\frac{1}{p_{j}}\mathrm{Tr}_{B}(\mathbb{I}_{A}\otimes\Pi_{j}^{B}\rho_{AB}\mathbb{I}_{A}\otimes\Pi_{j}^{B}).  
\end{equation*}
In the above equations, $\left\lbrace\Pi_{j}^{B}\right\rbrace$ is a complete set of von Neumann's measurements on subsystem $B$ satisfying the relations $\sum_{j}\Pi_{j}^{B}=\mathbb{I}_{B}$ and $\Pi_{j}^{B}\Pi_{k}^{B}=\delta_{jk}\Pi_{j}^{B}$.

The average quantum conditional entropy of subsystem $A$, given the measurement of the observable $O_{B}$ on subsystem $B$, is given by
\begin{equation*}
 S(\rho_{A|B})=\sum_{j}p_{j}S(\rho_{j}^{A}).  
\end{equation*}
Thus, a quantum extension for $J(A\text{:}B)$ can be defined as
\begin{equation}
\mathcal{J}(\rho_{AB}):=S(\rho_{A})-S(\rho_{A|B}).
\label{eq:QMI-J}
\end{equation}

Equations (\ref{eq:QMI-I}) and (\ref{eq:QMI-J}) for the quantum mutual information are not, in general, equivalent. The difference between them,
\begin{equation}
\mathcal{D}_{B}(\rho_{AB}):=\mathcal{I}(\rho_{AB})-\max_{O_{B}}\mathcal{J}(\rho_{AB}),
\label{eq:Discord-B}
\end{equation}
was named quantum discord \cite{Ollivier-Zurek} and is a measure of quantum correlation in bipartite systems. The maximum in Eq. (\ref{eq:Discord-B}) is obtained through the measurements (on subsystem $B$) that provide the maximal information about $A$. It is worthwhile mentioning that another version for QD, $\mathcal{D}_{A}$, can be obtained if we measure an observable $O_{A}$ on subsystem $A$. QD is an asymmetric quantity, in the sense that, in general, $\mathcal{D}_{A}\ne\mathcal{D}_{B}$, and has several interesting physical and information-theoretical interpretations \cite{Celeri-Review,Modi-Review}. 

We also observe that if we use the mutual information as a measure of total correlations, one can verify that QD may be written as the difference between the mutual information of subsystem $AB$ before and after a complete map of von Neumann's measurements is applied to one of the subsystems:
\begin{equation*}
\mathcal{D}_{B}(\rho_{AB})=\mathcal{I}(\rho_{AB})-\max_{\Pi^{B}}\mathcal{I}(\Pi^{B}(\rho_{AB})),
\end{equation*}
where $\Pi^{B}(\rho_{AB})=\sum_{j}(\mathbb{I}_{A}\otimes\Pi_{j}^{B})\rho_{AB}(\mathbb{I}_{A}\otimes\Pi_{j}^{B})$. In this alternative definition, QD can be interpreted as a measure of those correlations that are inevitably destroyed in the measurement process. 

\subsection{NMR Systems in Quantum Information Science}

The actual relation between non-classical correlations and the advantages obtained in quantum information science (compared with classical protocols) is an important issue that must be thoroughly analysed in seeking for a better understanding about the source(s) of the quantum speedup. Nuclear Magnetic Resonance (NMR) has been one of the leading experimental platforms for the implementation of protocols and algorithms in quantum information science (QIS) \cite{jones2011} and also for the simulation of quantum systems \cite{Peng-Suter}. In NMR implementations, the qubits are encoded in nuclear spins that are manipulated through carefully designed sequences of radio-frequency pulses. The information about the system state is obtained directly from the transverse magnetizations, that are the natural observables in NMR experiments \cite{oliveira2007}. The NMR density operator is typically highly mixed. In fact, Braunstein and co-authors showed that no entanglement was generated in most NMR implementations of QIS protocols \cite{Braunstein}. Moreover, Vidal proved that a large amount of entanglement must be generated if a pure-state quantum computation is to provide an exponential speedup in information processing \cite{Vidal}. The conjunction of these results led to a questioning about the possible classical nature of the NMR system and its QIS implementations. On the other hand, the generation of entanglement was shown to be a necessary but not a sufficient condition for the existence of a gain, even in pure-state quantum computation \cite{Jozsa.Linden}. Moreover, the source of quantum speedup in the mixed-state quantum computation scenario is even more subtle \cite{Linden.Popescu,Vedral-FP,Datta.PRL,Datta.IJQI}. In this case, QD seems to be the figure of merit for quantum advantage \cite{Eastin,Diogo-MZ}, although a definitive proof is still lacking. In this review we have a more modest goal. We deal only with the theoretical and experimental quantification and identification of quantum correlations in NMR systems at room temperature. We show the existence of quantumness in the correlations of highly mixed NMR states and study its peculiar dynamics under decoherence.  
 
\section{Quantification and Identification of Quantum Correlations}

\subsection{Symmetric Quantum Discord}

As mentioned in the previous section, the original definition for QD is asymmetric with relation to the subsystem we choose to measure. To illustrate this fact, let us consider the following bipartite state
\begin{equation}
\rho_{AB}^{cq}=\sum_{i}p_{i}|\psi_{i}^{A}\rangle\langle\psi_{i}^{A}|\otimes\rho_{i}^{B},
\label{eq:CQ-states}
\end{equation}
where $\{p_{i}\}$ is a probability distribution, $\{|\psi_{i}^{A}\rangle\}\in\mathcal{H}_{A}$ is an orthonormal basis for subsystem $A$, and $\rho_{i}^{B}\in\mathcal{H}_{B}$ the reduced density operator of subsystem $B$. In this case, the QD obtained by measuring subsystem $A$, $\mathcal{D}_{A}(\rho_{AB}^{cq})$, vanishes while its alternative definition, $\mathcal{D}_{B}(\rho_{AB}^{cq})$, is null if and only if all $\rho_{i}^{B}$ commute, that is, if $[\rho_{i}^{B},\rho_{j}^{B}]=0$ for all $i$ and $j$. Moreover, numerical calculations indicate that whenever there exists an asymmetry in the state $\rho_{AB}$ regarding the exchange between the subsystems or, more specifically, whenever $S(\rho_{A})\ne S(\rho_{B})$ it will result that $\mathcal{D}_{A}(\rho_{AB})\ne\mathcal{D}_{B}(\rho_{AB})$ \cite{Maziero-SQD}. This asymmetry of quantum discord may seem weird in principle because one usually think about correlations as the information shared between subsystems. We note that such interpretation still holds if one recognizes that the degree of quantumness of the subsystems shall limit the amount of information that the observers can extract via local measurements.

The class of states in Eq. (\ref{eq:CC-states}) is the only one for which $\mathcal{D}_{A}(\rho_{AB}^{cc})=\mathcal{D}_{B}(\rho_{AB}^{cc})=0$. A quantity that indicates and, in some sense, quantifies the quantumness of correlations in states that cannot be cast as $\rho_{AB}^{cc}$ can be defined as \cite{Maziero-SQD}
\begin{equation}
\mathcal{D}(\rho_{AB}):=\mathcal{I}(\rho_{AB})-\max_{\Pi^{AB}}\mathcal{I}(\Pi^{AB}(\rho_{AB})),
\label{eq:Discord-AB}
\end{equation}
where the complete map of local von Neumann's measurements reads
\begin{equation*}
\Pi^{AB}(\rho_{AB}):=\sum_{j,k}(\Pi_{j}^{A}\otimes\Pi_{k}^{B})\rho_{AB}(\Pi_{j}^{A}\otimes\Pi_{k}^{B}),
\end{equation*}
with $\sum_{j}\Pi_{j}^{s}=\mathbb{I}_{s}$ and $\Pi_{j}^{s}\Pi_{k}^{s}=\delta_{jk}\Pi_{j}^{s}$ for $s=A\mbox{, }B$. This quantum correlation quantifier can be regarded as a symmetric version of QD. Once state $\Pi^{AB}(\rho_{AB})$ is a classical one, we define the classical counterpart of $\mathcal{D}$,
\begin{equation}
\mathcal{C}(\rho_{AB}):=\max_{\Pi^{AB}}\mathcal{I}(\Pi^{AB}(\rho_{AB})),
\label{eq:CC-AB}
\end{equation}
as a measure of the classical correlations in $\rho_{AB}$.

\subsection{Symmetric Quantum Discord for the Deviation Matrix}

In this subsection we show how to obtain the classical correlation (\ref{eq:CC-AB}) and the symmetric quantum discord (\ref{eq:Discord-AB}) for systems described by density matrices of the form \cite{Diogo-Quadrupolar}
\begin{equation}
\rho_{AB}=\frac{\mathbb{I}_{AB}}{4}+\epsilon\Delta\rho_{AB},
\label{deviation-m}
\end{equation}
with $\epsilon\ll1$ and $\Delta\rho_{AB}$ being the traceless deviation matrix, $\mathrm{Tr}(\Delta\rho_{AB})=0$ (Eq. (\ref{deviation-m}) is the typical density operator describing the state of NMR systems). In order to calculate the quantum mutual information we need to compute the von Neumann's entropy. For that purpose, let us use the eigen-decomposition of $\rho_{AB}$ to write
\begin{equation*}
\ln(\rho_{AB})=\sum_{j}\ln(1/4+\epsilon\lambda_{j})|\lambda_{j}\rangle\langle\lambda_{j}|.
\end{equation*}
By means of a Taylor expansion of $\ln(1/4+\epsilon\lambda_{j})$ we obtain
\begin{eqnarray}
S(\rho_{AB}) & = & -\frac{1}{\ln2}\mathrm{Tr}(\rho_{AB}\ln\rho_{AB})\nonumber \\
 & = & \mathrm{Tr}\left(\frac{\mathbb{I}_{AB}}{2}+\frac{\epsilon(2\ln2-1)}{\ln2}\Delta\rho_{AB}\right. \nonumber\\
& &  \hspace{3.5cm}  \left.-\frac{2\epsilon^{2}}{\ln2}\Delta\rho_{AB}^{2}+\cdots\right)\nonumber \\
 & = & 2-\frac{2\epsilon^{2}}{\ln2}\mathrm{Tr}(\Delta\rho_{AB}^{2})+\cdots.
\label{eq:Sab-dev-mat}
\end{eqnarray}

Once the reduced density matrices have the form
\begin{equation*}
\rho_{s}=\frac{\mathbb{I}_{s}}{2}+\epsilon\Delta\rho_{s}, 
\end{equation*}
with $s=A\mbox{, }B$, the same procedure can be used to compute
\begin{equation}
S(\rho_{s})=1-\frac{\epsilon^{2}}{\ln2}\mathrm{Tr}(\Delta\rho_{s}^{2})+\cdots.
\label{eq:Ss-dev-mat}
\end{equation}

Using (\ref{eq:Sab-dev-mat}) and (\ref{eq:Ss-dev-mat}) it is possible to write the quantum mutual information in terms of the deviation matrix, up to second order in $\epsilon$, as
\begin{equation}
\mathcal{I}(\rho_{AB})\approx\frac{\epsilon^{2}}{\ln2}[2\mathrm{Tr}(\Delta\rho_{AB}^{2})-\mathrm{Tr}(\Delta\rho_{A}^{2})-\mathrm{Tr}(\Delta\rho_{B}^{2})].
\label{eq:MI-dev-mat}
\end{equation}

The measured state obtained from $\rho_{AB}$ through a complete set of local-projective measurements reads
\begin{equation*}
\Pi^{AB}(\rho_{AB})=\frac{\mathbb{I}_{AB}}{4}+\epsilon\Delta\eta_{AB},
\end{equation*}
with the measured deviation matrix given by
\begin{eqnarray*}
\Delta\eta_{AB} & := & \Pi^{AB}(\Delta\rho_{AB})\nonumber \\
 & = & \sum_{j,k}(\Pi_{j}^{A}\otimes\Pi_{k}^{B})\Delta\rho_{AB}(\Pi_{j}^{A}\otimes\Pi_{k}^{B})
\end{eqnarray*}
and $\mathrm{Tr}\left[\Pi^{AB}(\Delta\rho_{AB})\right]=0$. Thus, the same procedure utilized when obtaining (\ref{eq:MI-dev-mat}) leads to the following expression for the mutual information of the measured state
\begin{equation}
\mathcal{I}(\Pi^{AB}(\rho_{AB}))\approx\frac{\epsilon^{2}}{\ln2}[2\mathrm{Tr}(\Delta\eta_{AB}^{2})-\mathrm{Tr}(\Delta\eta_{A}^{2})-\mathrm{Tr}(\Delta\eta_{B}^{2})],
\label{eq:MI-mes-dev-mat}
\end{equation}
where $\Delta\eta_{A(B)}=\mathrm{Tr}_{B(A)}(\Delta\eta_{AB})$. The symmetric QD and the classical correlation are then computed as shown in Eqs. (\ref{eq:Discord-AB}) and (\ref{eq:CC-AB}), respectively.

\subsection{Symmetric Quantum Discord for Two-Qubit States}

Any two-qubit state can be brought, through local unitary transformations, to the following form
\begin{eqnarray}
\rho_{AB}=&&\frac{1}{4}\left[\mathbb{I}_{AB}+\sum_{j=1}^{3}(a_{j}\sigma_{j}^{A}\otimes\mathbb{I}_{B}+b_{j}\mathbb{I}_{A}\otimes\sigma_{j}^{B}) \right. \nonumber \\
& & \hspace{3.5cm} \left.+\sum_{j=1}^{3}c_{j}\sigma_{j}^{A}\otimes\sigma_{j}^{B}\right],
\label{eq:2qubits-states}
\end{eqnarray}
where $\{\sigma_{j}^{s}\}$ are the Pauli operators acting on $\mathcal{H}_{s}$ and $a_{j},b_{j},c_{j}\in\mathbb{R}$. The measured state, modulo local unitary transformations, can be written as
\begin{equation*}
\Pi^{AB}(\rho_{AB})=\frac{1}{4}(\mathbb{I}_{AB}+\alpha\sigma_{3}^{A}\otimes\mathbb{I}_{B}+\beta\mathbb{I}_{A}\otimes\sigma_{3}^{B} +\gamma\sigma_{3}^{A}\otimes\sigma_{3}^{B}),
\end{equation*}
with 
\begin{equation*}
\alpha:=\sum_{j}a_{j}z_{j}^{A}\mathtt{, }\hspace{0.2cm} \beta:=\sum_{j}b_{j}z_{j}^{B}\mathtt{, }\hspace{0.2cm} \gamma:=\sum_{j}c_{j}z_{j}^{A}z_{j}^{B}.  
\end{equation*}
The parameters
\begin{eqnarray*}
z_{1}^{s}&=&2\sin\theta_{s}\cos\theta_{s}\cos\phi_{s}, \nonumber \\
z_{2}^{s}&=&2\sin\theta_{s}\cos\theta_{s}\sin\phi_{s}, \nonumber \\
z_{3}^{s}&=&2\cos^{2}\theta_{s}-1,
\end{eqnarray*}
with 
\begin{equation*}
\theta_{s}\in[0,\pi/2] \text{ and } \phi_{s}\in[0,2\pi],  
\end{equation*}
determine the measurement direction for the subsystem $s=A\mbox{, }B$. It seems that the maximization problem for obtaining the classical correlation cannot be solved in general. Nevertheless, for Bell-diagonal states,
\begin{equation}
\rho_{AB}^{bd}=\frac{1}{4}\left(\mathbb{I}_{AB}+\sum_{j=1}^{3}c_{j}\sigma_{j}^{A}\otimes\sigma_{j}^{B}\right),\label{eq:Bell-D}
\end{equation}
the correlations can be obtained analytically. In this case we have that $S(\rho_{A}^{bd})=S(\rho_{B}^{bd})=1$ and thus Eq.(\ref{eq:CC-AB}) become
\begin{equation*}
C(\rho_{AB}^{bd})=2-\min_{\gamma}S\left[\Pi^{AB}(\rho_{AB}^{bd})\right], 
\end{equation*}
with
\begin{equation*}
S(\Pi^{AB}(\rho_{AB}^{bd}))=-\sum_{j=1}^{4}\frac{1+(-1)^{j}\gamma}{4}\log_{2}\frac{1+(-1)^{j}\gamma}{4}.
\end{equation*}
We note that the minimum of the measured state entropy is obtained by maximizing $|\gamma|$. It results that 
\begin{equation*}
|\gamma|\le\kappa:=\max(|c_{1}|,|c_{2}|,|c_{3}|). 
\end{equation*}
Thus, an analytical expression for the symmetric QD for Bell-Diagonal states is obtained as \cite{Maziero-SQD}:
\begin{equation}
\mathcal{D}(\rho_{AB}^{bd})=\sum_{j,k=0}^{1}\lambda_{jk}^{bd}\log_{2}4\lambda_{jk}^{bd}-\mathcal{C}(\rho_{AB}^{bd}),
\label{eq:SQD-ANA-BD}
\end{equation}
with
\begin{equation}
\mathcal{C}(\rho_{AB}^{bd})=\frac{1}{2}\sum_{j=0}^{1}[1+(-1)^{j}\kappa]\log_{2}[1+(-1)^{j}\kappa]
\label{eq:SCC-ANA-BD}
\end{equation}
and $\lambda_{jk}^{bd}=[1+(-1)^{j}c_{1}-(-1)^{j+k}c_{2}+(-1)^{j}c_{3}]/4.$ 

\subsection{Peculiar Dynamics of Correlations under Decoherence\label{SC-theory}}

\begin{figure}
\begin{centering}
\includegraphics[width=0.45\textwidth]{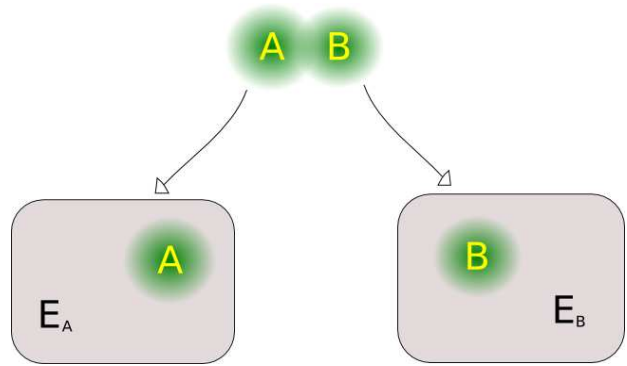}
\end{centering}
\caption{Schematic representation of the two systems $A$ and $B$, initially prepared in a state $\rho_{AB}$, evolving under the action of local independent phase channels.}
\label{AEaBEb}
\end{figure}

We shall use the analytical formulas for the symmetric QD (\ref{eq:SQD-ANA-BD}) and for its classical counterpart (\ref{eq:SCC-ANA-BD}) in Bell-diagonal states to study the dynamics of these correlations under the influence of local independent Markovian environments that inject phase noise into the system, as schematically depicted in Fig. \ref{AEaBEb} \cite{Maziero-SC}. Under such conditions, the two subsystems $A$ and $B$, that are initially prepared in a Bell-diagonal state (\ref{eq:Bell-D}), evolve via the following quantum operation \cite{Nielsen-Chuang} 
\begin{eqnarray}
\rho_{AB}^{bd}(p) & = & \mathcal{E}_{PD}(\rho_{AB}^{bd})\nonumber \\
 & = & \sum_{j,k}(K_{j}^{A}\otimes K_{k}^{B})\rho_{AB}^{bd}(K_{j}^{A}\otimes K_{k}^{B})^{\dagger}
 \label{eq:sum-op-rep}\\
 & = & \frac{1}{4}\begin{bmatrix}1+\gamma & 0 & 0 & \alpha-\beta\\
0 & 1-\gamma & \alpha+\beta & 0\\
0 & \alpha+\beta & 1-\gamma & 0\\
\alpha-\beta & 0 & 0 & 1+\gamma
\end{bmatrix}.
\label{eq:evolved-BD-PD}
\end{eqnarray}
All matrices of the article are represented in the standard computational basis $\{|00\rangle,|01\rangle,|10\rangle,|11\rangle\}$, where $|ij\rangle:=|i\rangle\otimes|j\rangle$ and $\{|0\rangle,|1\rangle\}$ are the eigenstates of the Pauli matrix $\sigma_{z}$. The Kraus' operators appearing in the operator-sum representation (\ref{eq:sum-op-rep}) for the phase-damping channel are
\begin{eqnarray}
K_{0}^{s} & = & \sqrt{1-\frac{p_{s}}{2}}\begin{bmatrix}1 & 0\\0 & 1 \end{bmatrix},\nonumber \\
K_{1}^{s} & = & \sqrt{\frac{p_{s}}{2}}\begin{bmatrix}1 & 0\\0 & -1\end{bmatrix},
\label{phase}
\end{eqnarray}
with $s=A\mbox{, }B$ and $p_{A}=p_{B}:=p$ being the parametrized times. This means that we considered identical environments for both subsystems. Besides, we have defined
\begin{eqnarray*}
\alpha & := & (1-p)^{2}c_{1},\nonumber \\
\beta & := & (1-p)^{2}c_{2},\nonumber \\
\gamma & := & c_{3}.
\end{eqnarray*}

We see that the evolved density matrix (\ref{eq:evolved-BD-PD}) is in Bell-diagonal form (\ref{eq:Bell-D}). Thus, the symmetric QD $\mathcal{D}\left[\rho_{AB}^{bd}(p)\right]$ and the classical correlation $\mathcal{C}\left[\rho_{AB}^{bd}(p)\right]$ are given by (\ref{eq:SQD-ANA-BD}) and (\ref{eq:SCC-ANA-BD}), respectively, with 
\begin{equation*}
\kappa=\max(|\alpha|,|\beta|,|\gamma|) 
\end{equation*}
and
\begin{equation*}
\lambda_{jk}^{bd}=[1+(-1)^{j}\alpha-(-1)^{j+k}\beta+(-1)^{j}\gamma]/4. 
\end{equation*}

\begin{flushleft}
\begin{figure}
\begin{centering}
\includegraphics[scale=0.88]{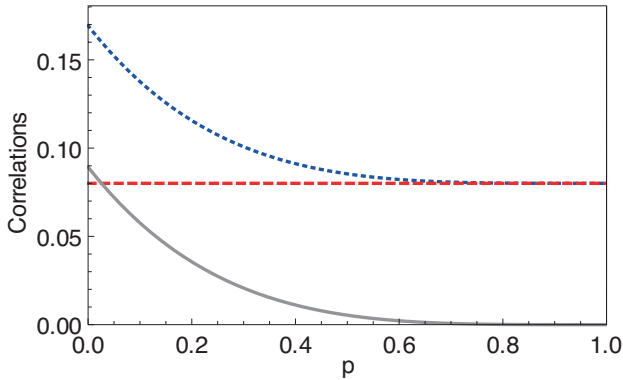}
\par\end{centering}
\caption{Total correlation (blue dotted line), classical correlation (red dashed line), and quantum discord (gray continuous line) for a Bell-diagonal initial state evolving under local independent phase-damping channels. For this figure, the parameters in the initial state (\ref{eq:Bell-D}) were chosen as: $c_{1}=0.06$, $c_{2}=0.3$, and $c_{3}=0.33$. For this state the classical correlation is not affected by the environment while the quantum correlation decays monotonically.}
\label{teo-SC-Cconst}
\end{figure}
\par\end{flushleft}

From the analytical results we readily identify three general and different kinds of dynamic behaviour of the correlations under decoherence:
\begin{description}
\item (i) If $|c_{3}|\ge|c_{1}|,|c_{2}|$ we have $\kappa=|c_{3}|$ and the classical correlation does not depend on the parametrized time $p$. As the quantum correlation decays monotonically going to zero in the asymptotic state limit, the classical correlation is equal to the mutual information of this state, i.e.
\begin{equation*}
\mathcal{C}\left[\rho_{AB}^{bd}(p)\right]=\mathcal{C}\left[\rho_{AB}^{bd}(p=0)\right]=\mathcal{I}\left[\rho_{AB}^{bd}(p=1)\right].
\end{equation*}
This kind of dynamical behaviour of the correlations is exemplified in Fig. \ref{teo-SC-Cconst} for the initial state $\rho_{AB}^{bd}$ with $c_{1}=0.06$, $c_{2}=0.3$, and $c_{3}=0.33$.

\begin{center}
\begin{figure}
\begin{centering}
\includegraphics[scale=0.88]{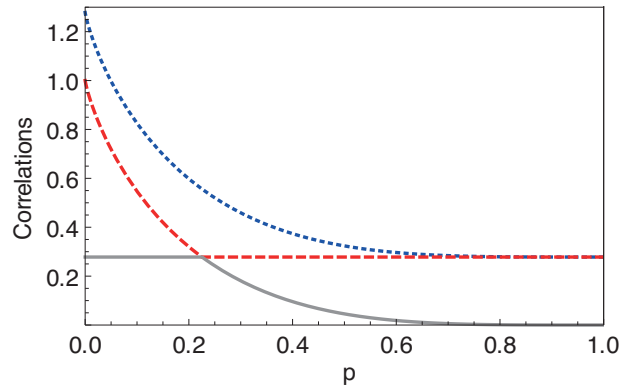}
\par\end{centering}
\caption{Total correlation (blue dotted line), classical correlation (red dashed line), and quantum discord (gray continuous line) for the a Bell-diagonal initial state evolving under local independent phase-damping channels. For this figure, the parameters in the initial state (\ref{eq:Bell-D}) were chosen as: $c_{1}=1$, $c_{2}=-0.6$, and $c_{3}=0.6$. For this state the sudden change occurs at $p_{sc}\approx0.22$ and the quantum discord (classical correlation) remains constant (decays monotonically) for $p\lesssim0.22$ with the opposite scenario taking place for $p\gtrsim0.22$.}
\label{teo-SC-SC}
\end{figure}
\par\end{center}

\item (ii) If $|c_{1}|\ge|c_{2}|,|c_{3}|$ or $|c_{2}|\ge|c_{1}|,|c_{3}|$ and $|c_{3}|\ne0$ we have a peculiar dynamics of the correlations under decoherence with a sudden change in behaviour. Classical correlation decays monotonically until a specific parametrized time
\begin{equation}
p_{sc}=1-\sqrt{\frac{|c_{3}|}{\max(|c_{1}|,|c_{2}|)}},
\end{equation}
and remains constant form this time afterwards. For $p<p_{sc}$ we have $\kappa=(1-p)^{2}\max(|c_{1}|,|c_{2}|)$ and $\mathcal{C}$ decays monotonically. On the other hand, for $p\ge p_{sc}$ we have $\kappa=|c_{3}|$ and $\mathcal{C}$ is constant:
\begin{equation*}
\mathcal{C}\left[\rho_{AB}^{bd}(p>p_{sc})\right]=\mathcal{I}\left[\rho_{AB}^{bd}(p=1)\right],
\end{equation*}
and the decay rate of $\mathcal{D}$ changes abruptly in $p=p_{sc}$. The dynamical behaviour of correlations with a sudden change in its decay rate is exemplified in Fig. \ref{teo-SC-SC} for the initial state $\rho_{AB}^{bd}$ with $c_{1}=1$, $c_{2}=-0.6$ and $c_{3}=0.6$. For this example, not only the classical correlation remains constant in a certain time interval but also the quantum correlation can be immune against decoherence. The regimes where $\mathcal{D}=\mathrm{constant}$ and $\mathcal{C}=\mathrm{constant}$ were named classical and quantum decoherence regimes in Ref. \cite{Mazzola}.

\begin{center}
\begin{figure}
\begin{centering}
\includegraphics[scale=0.88]{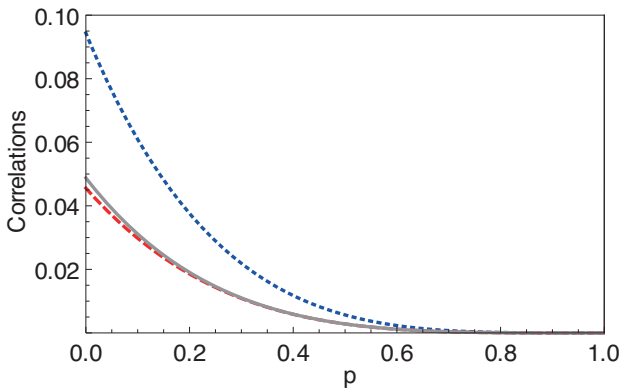}
\par\end{centering}
\caption{Total correlation (blue dotted line), classical correlation (red dashed line), and quantum discord (gray continuous line) for the a Bell-diagonal initial state evolving under local independent phase-damping channels. For this figure, the parameters in the initial state (\ref{eq:Bell-D}) were chosen as: $c_{1}=c_{2}=0.25$ and $c_{3}=0.0$. For this state all correlations decay monotonously with time.}
\label{teo-monotonic}
\end{figure}
\par\end{center}

\item (iii) Finally, if $c_{3}=0$ both correlations $\mathcal{C}$ and $\mathcal{D}$ decay monotonically. We exemplify this kind of dynamic behaviour in Fig. \ref{teo-monotonic} for a Bell-diagonal initial state with $c_{1}=c_{2}=0.25$ and $c_{3}=0.0$.

\end{description}

\subsection{Classicality Witness\label{sec:class-wit}}

The set composed by separable states of the form (\ref{eq:sep-rho}) is convex and hence, there exists a linear hermitian operator, called entanglement witness, that can be used to distinguish separable from entangled states. It is natural to ask whether the same approach can be utilized to identify classical states. It turns out that the set formed by states with null QD is not convex and, as a consequence, linear witnesses cannot be applied any more. To verify this assertion, let us consider a linear hermitian operator $W_{l}$ whose mean value is non-negative for classical states, i.e.
\begin{equation*}
\langle W_{l}\rangle_{\rho_{AB}^{cc}}=\mathrm{Tr}(W_{l}\rho_{AB}^{cc})\ge0,
\end{equation*}
and $\langle W_{l}\rangle_{\rho_{AB}}<0$ for quantum-correlated states. Thus we note that for the separable states (\ref{eq:sep-rho}),
\begin{equation*}
\langle W_{l}\rangle_{\rho_{AB}^{sep}}=\sum_{j}p_{j}\langle W_{l}\rangle_{\rho_{j}^{A}\otimes\rho_{j}^{B}}\ge0.
\end{equation*}
Therefore linear witnesses cannot be applied to identify separable discordant states. Next we introduce a non-linear classicality witness providing a sufficient condition for absence of quantumness in the correlations of two-qubit states of the form (\ref{eq:2qubits-states}) \cite{Maziero-W}. First we consider the following set of hermitian operators
\begin{eqnarray*}
O_{j} & = & \sigma_{j}^{A}\otimes\sigma_{j}^{B},\nonumber \\
O_{4} & = & \vec{z}\cdot\vec{\sigma}_{A}\otimes\mathbb{I}_{B}+\mathbb{I}_{A}\otimes\vec{w}\cdot\vec{\sigma}_{B},
\end{eqnarray*}
where $j=1,2,3$ and $\vec{z},\vec{w}\in\mathbb{R}^{3}$ ($|\vec{z}|=|\vec{w}|=1$) should be randomly chosen. Let us consider a relation among the averages of these operators as follows
\begin{equation}
W_{\rho_{AB}}=\sum_{i=1}^{3}\sum_{j=1+1}^{4}|\langle O_{i}\rangle_{\rho_{AB}}\langle O_{j}\rangle_{\rho_{AB}}|,\label{eq:class-witness}
\end{equation}
where $|x|$ is the absolute value of $x$. We see that $W_{\rho_{AB}}=0$ if and only if the mean value of at least three of the four observables defined above are zero. As
\begin{eqnarray*}
\langle O_{i}\rangle_{\rho_{AB}} & = & c_{i}\mbox{ for }i=1,2,3,\nonumber \\
\langle O_{4}\rangle_{\rho_{AB}} & = & \vec{z}\cdot\vec{a}+\vec{w}\cdot\vec{b},
\end{eqnarray*}
with $\vec{a}=(a_{1},a_{2},a_{3})$ and $\vec{b}=(b_{1},b_{2},b_{3})$, the only way to warrant that $W_{\rho_{AB}}=0$ is if $\rho_{AB}$ is of the type
\begin{eqnarray*}
\chi_{i} & = & \frac{1}{4}(\mathbb{I}_{AB}+c_{i}\sigma_{i}^{A}\otimes\sigma_{i}^{B})\mbox{ for }i=1 \text{ or } i=2 \text{ or } i=3,\nonumber \\
\chi_{4} & = & \frac{1}{4}\left[\mathbb{I}_{AB}+\sum_{j=1}^{3}(a_{j}\sigma_{j}^{A}\otimes\mathbb{I}_{B}+b_{j}\mathbb{I}_{A}\otimes\sigma_{j}^{B})\right].
\end{eqnarray*}
It turns out that all theses states can be cast as those in Eq. (\ref{eq:CC-states}) and, therefore, do not possess quantum correlations. As a consequence, it follows that $W_{\rho_{AB}}=0$ is a sufficient condition for $\rho_{AB}$ to be classically correlated or to have no correlations at all.

For Bell-diagonal states, $W_{\rho_{AB}}=0$ is also a necessary condition for the absence of quantum discord in the system. This result is obtained by noting that $\rho_{AB}^{bd}$ being classically correlated implies that it must have the form $(\mathbb{I}_{AB}+c_{i}\sigma_{i}^{A}\otimes\sigma_{i}^{B})/4$ with $i=1$ or $i=2$ or $i=3$, implying that $W_{\rho_{AB}}=0$.

For the experimental implementation of the classicality witness (\ref{eq:class-witness}) in the NMR context\footnote{A modified version of this classicality witness was implemented in the optical context \cite{Aguilar}}, that will be presented bellow, it is useful to rewrite the witness $W_{\rho_{AB}}$ in terms of the qubits magnetizations, that are the natural observables accessed in NMR experiments. First we observe that
\begin{equation*}
\langle O_{i}\rangle_{\rho_{AB}}=\langle\sigma_{1}^{A}\otimes\mathbb{I}_{B}\rangle_{\xi_{i}}
\end{equation*}
with
\begin{equation*}
\xi_{i}=U{}_{A\rightarrow B}[R_{n_{i}}(\theta_{i})\rho_{AB}R_{n_{i}}^{\dagger}(\theta_{i})]U{}_{A\rightarrow B},
\end{equation*}
where 
\begin{equation*}
U_{A\rightarrow B}=|0\rangle\langle0|\otimes\mathbb{I}_{B}+|1\rangle\langle1|\otimes\sigma_{1}^{B}
\end{equation*}
is the controlled-NOT gate with qubit $A$ as control and $R_{n_{i}}(\theta_{i})=R_{n_{i}}^{A}(\theta_{i})\otimes R_{n_{i}}^{B}(\theta_{i})$ with $R_{n_{i}}^{s}(\theta_{i})$ being a local rotation on qubit $s=A\mbox{, }B$ by an angle $\theta_{i}$ ($\theta_{1}=0\mbox{, }\theta_{2}=\theta_{3}=\pi/2$) in direction $\hat{n}_{i}$ ($\hat{n}_{2}=\hat{y}\mbox{, }\hat{n}_{3}=\hat{z}$).

\section{Quantum Discord in NMR Systems}

\subsection{The System\label{sec:levelC1}}

In order to experimentally follow the dynamic of correlations under decoherence and to implement the classicality witness discussed in the preceding section, it is necessary to prepare, manipulate, and measure the state of a two-qubit system in the laboratory.  In the NMR scenario, two-qubit systems can be achieved using samples presenting either two J coupled nuclear spins-$1/2$ (where J means scalar spin-spin coupling) or one quadrupolar spin-$3/2$ system in the presence of a local electric field gradient (named as quadrupolar spin systems). There are many examples of two dipolar J coupled spins-$1/2$ in NMR, for instance, $^{1}\textnormal{H}$  and $^{13}\textnormal{C}$ nuclei in chloroform (CHCl$_{3}$) or $^{1}\textnormal{H}$ and $^{31}\textnormal{P}$ nuclei in phosphoric acid (H$_{3}$PO$_{4}$) \cite{jones2011}. Two-qubit quadrupolar systems used so far in QIP are usually comprised by spin-$3/2$ nuclei in single  crystals \cite{kampermann2005} or in lyotropic liquid crystals \cite{oliveira2007,sinha2001}. $^{23}$Na and $^{7}$Li nuclei in lyotropic liquid crystals based on Sodium Dodecyl Sulfate (SDS) \cite{oliveira2007} and Lithium Tetrafluoroborate (LiBF$_{4}$) \cite{sinha2001}, as well as $^{23}$Na in Sodium Nitrate (NaNO$_{3}$) single crystals \cite{kampermann2005} are some examples of these quadrupolar systems.

All spin systems cited above are equally good representations of two-qubit sytems. However, the nuclear spin interactions that drive their quantum evolution are distinct, leading to significant differences in the techniques used for state preparation, manipulation and read-out. Besides, as a result of the specific characteristics of each spin interaction, there are unique features in their decoherence behaviour. Hence, in the next sections we shall discuss each system separately.  

\begin{figure}[h!]
\includegraphics[width=0.47\textwidth]{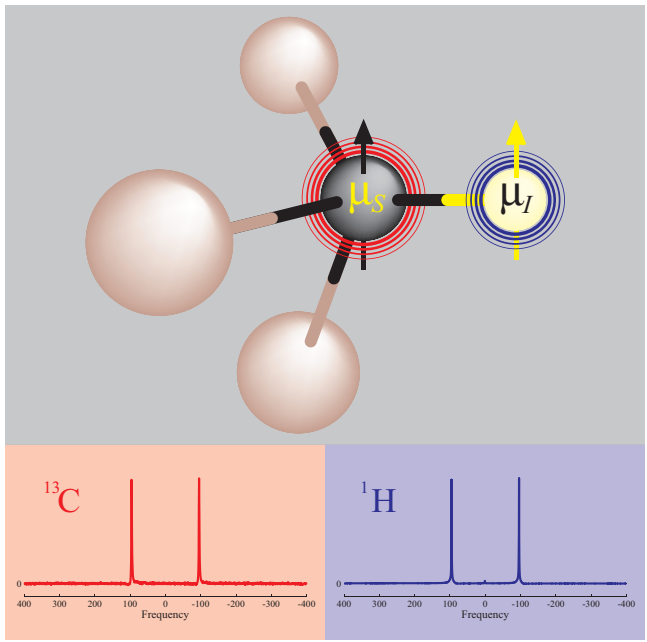}
\caption{Top: Schematic representation of Chloroform molecule (HCCl$_{3}$). The red (blue) circles depict perturbation and response of Carbon (Hydrogen) nucleus at its respective resonance frequency. Bottom: The normalized equilibrium spectra of Carbon (Red) and Hydrogen (Blue) nuclei.}
\label{fig:MoleculaCloroformio}
\end{figure}

\subsubsection{Two Dipolar J Coupled Nuclear Spin-$1/2$ Systems - Liquid Sample\label{sec:levelC1A}}

A schematic representation of a $^{13}\textnormal{C}$ enriched chloroform molecule (CHCl$_{3}$) is depicted in Fig. \ref{fig:MoleculaCloroformio}. As already mentioned, the two-qubits are comprised by the nuclear $^{1}\textnormal{H}-^{13}\textnormal{C}$ spin pair in the molecule. However, due to sensitivity reasons, one need to have many replicas of these two-qubit units, i.e, the system need to be constituted by an ensemble of spin pairs where each spin in the pair interacts only with its counterpart. This is achieved by diluting the $^{13}\textnormal{C}$ enriched CHCl$_{3}$ molecules in a deuterated solvent. Since the deuteration and the low natural abundance of $^{13}\textnormal{C}$ make the solvent molecules magnetically inert, in the NMR point of view this system can be well described as an ensemble of isolated spin pairs, i.e, many replicas of two-qubit units. Thus, all the results reported here were obtained using a sample prepared by dissolving $100$ mg of $99\%$ $^{13}$C-labeled CHCl$_{3}$ in $0.2$ ml of $99.8\%$ Acetone-$d6$, and placing this in a Wildmad LabGlass $5$ mm tube. Both samples were provided by the Cambridge Isotope Laboratories - Inc. NMR experiments were performed at $25^{\circ}$ C using a Varian $500$ MHz Spectrometer and a 5-mm double-resonance probe tuned to 1H and 13C nuclei. 

In NMR experiments the nuclear spins are placed in a strong static magnetic field $B_{0}$, whose direction is taken to be the positive $z$. Spins manipulation is achieved by applying time dependent radio-frequency (rf) fields on both spins, which interact to each other and with their local environments. Thus, in a double rotating frame with rf frequencies $\omega_{rf}^{H}$ and $\omega_{rf}^{C}$, the Hamiltonian of a two coupled spins is represented by
\begin{eqnarray}
\mathcal{H}=&&-(\omega_{H}-\omega _{rf}^{H})\mathbf{I}_{z}^{H}-(\omega_{C}-\omega _{rf}^{C})\mathbf{I}_{z}^{C}+2\pi J\mathbf{I}_{z}^{H}\mathbf{I}_{z}^{C} \nonumber \\
&&+\omega_{1}^{H}\left(\mathbf{I}_{x}^{H}\cos \varphi^{H}+\mathbf{I}_{y}^{H}\sin \varphi^{H}\right) \nonumber \\
&&+\omega_{1}^{C}\left(\mathbf{I}_{x}^{C}\cos \varphi^{C}+\mathbf{I}_{y}^{C}\sin \varphi^{C}\right)+\mathcal{H}_{Env}(t), 
\label{HamiltonianoLiquido}
\end{eqnarray}
where $\mathbf{I}_{u}^{H}\left(\mathbf{I}_{v}^{C}\right)$ is the spin angular momentum operator in the $u,v=x,y,z$ direction for $^{1}$H ($^{13}$C); $\varphi ^{H}\left(\varphi ^{C}\right)$ and $\omega _{1}^{H}\left(\omega _{1}^{C}\right)$ define phase and power, respectively, of the radio-frequency field for the $^{1}$H ($^{13}$C) nuclei. The first two terms describe the free precession of $^{1}$H and $^{13}$C nuclei about $B_{0}$, with Larmor frequencies
\begin{equation*}
\frac{\omega_{H}}{2\pi}\approx 500 \mbox{ MHz} \text{ and } \frac{\omega_{C}}{2\pi}\approx 125\mbox{ MHz}   
\end{equation*}
The third term describes the scalar spin-spin coupling (also refereed as $J$ coupling in the NMR literature) of frequency 
\begin{equation*}
J\approx 215.1\mbox{ Hz}.  
\end{equation*}
The fourth and fifth terms are the external rf fields applied to manipulate the $^{1}$H and $^{13}$C nuclear spins, respectively. $\mathcal{H}_{Env}(t)$ represents time dependent fields resulting from the random fluctuating in the interactions between the spins and their environment. This term leads to spin relaxation and decoherence and includes interactions with the chlorine nuclei as well as higher order terms in the spin-spin coupling.

The density operator for any quantum system in contact with a thermal bath can be represented as
\begin{equation}
\rho=\frac{\exp(-\beta\mathcal{H})}{\mathrm{Tr}\exp(-\beta\mathcal{H})}, 
\label{OperadorDensidade} 
\end{equation}
where $\beta=1/k_{B}T$ with $k_{B}$ being Boltzmann's constant and $T$ the absolute temperature. For two-qubit NMR systems at room temperature, the thermal energy is much greater than the magnetic energy:
\begin{equation*}
\varepsilon = \frac{\hbar \gamma B_{0}}{ 4 k_{B} T} \approx 10^{-5},  
\end{equation*}
where $\gamma$ is the gyromagnetic ratio. In this case, it is possible to write the density operator in the so called high-temperature expansion:
\begin{eqnarray}
\rho\approx\frac{\mathbb{I}^{ab}}{4}+\varepsilon\Delta\rho, 
\label{OperadorAproximado}
\end{eqnarray}
with $\Delta\rho$ being the deviation matrix. We observe that $\Delta\rho$ is the part of the system density matrix which is actually manipulated (via rf pulses) and accessed in NMR experiments. 

Two unitary transformations are frequently used in NMR experiments: radio-frequency pulses and free evolutions of the spin systems under the coupling and static fields. The rf pulses are characterized by some intrinsic parameters as amplitude (power), duration, frequency and phase. By proper set of these parameters, it is possible to induce nuclear spin rotations about any axis and with any rotation angle. For example, a non-selective $^{1}\textnormal{H}$ spin rotation (also known in the NMR jargon as hard pulses) of $\pi/2$ can be performed by a rf pulse with duration of $7.4$ $\mu$s. Using these time values we acquire the equilibrium spectrum (NMR spectrum acquired after applying the rf pulses to the thermal equilibrium state) for each nuclear species (see Fig. \ref{fig:MoleculaCloroformio}). Analogously, the free evolution corresponds to precession of the nuclear spins around $B_{0}$.

\begin{figure}[h!]
\includegraphics[width=0.47\textwidth]{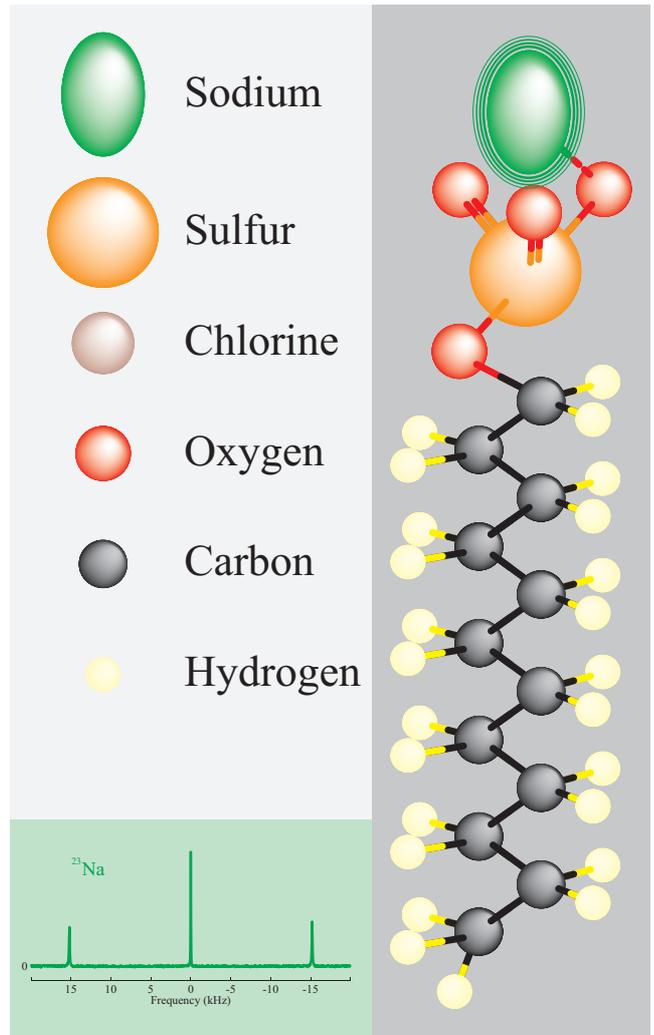}
\caption{Left-top: Labels of atomic compounds of the molecules. Left-bottom: The normalized equilibrium spectrum of the Sodium nucleus. Right: Cartoon of the Sodium Dodecil Sulfate molecule (SDS). The green ellipses depict the perturbation and response of Sodium nucleus at its respective resonance frequencies.}
\label{fig:MoleculaSDS}
\end{figure}

\subsubsection{Nuclear Quadrupolar System - Liquid Cristal Sample\label{sec:levelC1B}}

As already mentioned, $^{23}$Na ($I=3/2$) in a SDS molecule of a lyotropic liquid crystal is a good representation of a two-qubit quadrupolar spin system. As depicted in Fig. \ref{fig:MoleculaSDS}, this molecule has a chemical structure were the $^{23}$Na nucleus is far from other NMR active high abundant nuclei, mostly $^{1}$H, so the magnetic dipolar interaction of $^{23}\textnormal{Na}$ with other nuclei in the SDS molecule can be neglected. Besides, since the liquid crystal is prepared with heavy water ($\textnormal{D}_{2}\textnormal{O}$), the dipolar interaction with the $^{1}$H nuclear spins of water is also week (deuterium is a low $\gamma$ nucleus). Thus, in the SDS $\textnormal{D}_{2}\textnormal{O}$ based liquid crystal, the interaction between the $^{23}$Na quadrupole moment and the electric field gradient produced by the electrical charges in the vicinity (usually named as quadrupolar coupling) can be considered as the only internal spin interaction that affects the quantum system evolution. However, the local electric field gradient is defined by the charges configuration in the $^{23}$Na nucleus proximity and to produce a system were each $^{23}$Na environment is replicated along the sample, all SDS molecules needs to have similar orientations. Fortunately, the strong static magnetic field applied in NMR experiments induces the alignment of the SDS molecules in the lyotropic liquid crystal, naturally producing a situation were the local electric field gradients are identical at each $^{23}$Na site. Furthermore, the anysotropic internal motions of the SDS molecules reduces the strength of the quadrupolar coupling to smaller values, facilitating the manipulation of the spin system by rf pulses.  Under such conditions, in a rotating frame with frequency $\omega _{rf}$, the $^{23}$Na nuclear spin Hamiltonian can be described by:
\begin{eqnarray}
\mathcal{H} &=&-\left( \omega_{L}-\omega _{rf}\right) \mathbf{I}_{z}+ \frac{\omega_{Q}}{6}\left(  3\mathbf{I}_{z}^{2}-{\mathbf{I}}^{2}\right)  \nonumber \\
&&+\omega _{1}\left( \mathbf{I}_{x}\cos \varphi +\mathbf{I}_{y}\sin \varphi\right)  +\mathcal{H}_{Env}^{Q}(t) \text{,} \label{HamiltonianoCristal}
\end{eqnarray}
where $\omega_{Q}$ is the strength of the quadrupolar coupling in frequency units and $\omega_{L}$ is the Larmor frequency ($\left\vert \omega_{L}\right\vert \gg\left\vert \omega_{Q}\right\vert $). The spin angular momentum operators are represented by its $u$-component $\mathbf{I}_{u}$ ($u=x,y,z$) and its square modulus ${\mathbf{I}}^{2}$. The first term of Eq. (\ref{HamiltonianoCristal}) describes the Zeeman interaction between a static magnetic field and the nuclear spin, while the second term accounts for the static first order quadrupolar interaction. As in Eq. (\ref{HamiltonianoLiquido}), the third and fourth terms correspond to an externally applied rf field and the time dependent coupling of the $^{23}$Na nuclear spin with its environment, respectively. In this case, $\mathcal{H}_{Env}^{Q} (t)$ results mainly from the motion induced fluctuations in the local electric field gradients with smaller contributions from the weak dipolar interactions between the $^{23}$Na and other NMR active spins in the surroundings \cite{sarthour2003,slichter1992}.

The experimental results presented here were obtained using a liquid crystal sample prepared with $20.9\%$ of SDS ($95\%$ of purity), $3.7\%$ of decanol, and $75.4\%$ deuterium oxide, following the procedure described in Ref. \cite{radley1976}. The $^{23}$Na NMR experiments were performed in a $400$ MHz - VARIAN INOVA spectrometer using a $7$ mm solid-state NMR probe head at $26^{\circ}$C. In the left-bottom of Fig. \ref{fig:MoleculaSDS} we show the $^{23}$Na equilibrium spectrum in our SDS liquid crystal sample. The quadrupole coupling frequency is directly obtained from the separation between the central and the satellite lines \cite{jaccard1986,auccaise2008}, which in this particular case gives
\begin{equation*}
\nu_{Q}=\frac{\omega_{Q}}{2\pi} = 15\mbox{ kHz}. 
\end{equation*} 

Once the general features of the two main two-qubit NMR systems were presented, we now wish to discuss how these systems can be used for QIP. To do so, one needs to show how to perform the three main QIP steps, i.e state preparation, manipulation and read-out. This will be the focus of the next three subsections.   

\subsection{State Preparation\label{sec:levelC2}}

Most of the NMR QIP implementations relies on effective pure states, usually refereed as pseudo-pure states. Among others \cite{knill1998,cory1998,gershenfeld1997}, the main procedures for pseudo-pure state preparation are the so called spatial \cite{cory1998} and temporal averaging \cite{knill1998}, which we shall discuss in this section. 

The general idea behind the spatial averaging is, starting from the thermal equilibrium, to apply a set of rf and magnetic field gradient pulses to manipulate the spin populations and coherences (diagonal and off-diagonal elements of the density matrix in the computational basis, respectively) in order to produce the desired quantum state. While the rf pulse acts changing the density matrix populations and coherences through specific spin rotations, the field gradients are used to clean-up undesired coherences. Indeed, the application of the gradient pulses characterizes the spacial averaging, since they produce a magnetic field distribution along the sample that induces a random phase distribution in the individual quantum coherences, so they do not contribute to the final average state. For instance, the quantum state $\left|11\right\rangle$ of the computational basis can be prepared by applying the following pulse sequence to the thermal equilibrium state:
\begin{eqnarray*}
&\left(\frac{\pi}{2}\right)^{H,C}_{-x} \rightarrow \left(\frac{1}{4J}\right) \rightarrow \left(\frac{\pi}{2}\right)^{H,C}_{y} \rightarrow \left(\frac{1}{4J}\right) \rightarrow \left(\frac{\pi}{2}\right)^{H,C}_{-x} \rightarrow \\
& \left\{G_{z}\right\} \rightarrow \left(\frac{\pi}{4}\right)^{H,C}_{-y} \rightarrow \left(\frac{1}{2J}\right) \rightarrow \left(\frac{\pi}{6}\right)^{H,C}_{x} \rightarrow \left\{G_{z}\right\},
\end{eqnarray*}   
where $\left(\theta\right)^{H,C}_{d}$ is a rotation by an angle $\theta$ in $d$ direction applied simultaneously on $^{1}$H and $^{13}$C nuclear spins. $\left(1/nJ\right)$ represents the duration of a free evolution. Last $\left\{G_{z}\right\}$ states for a magnetic field gradient in $z$ direction. Note that, in this case, the specific rotations are implemented by hard rf pulses with proper phases (sub indices) and amplitudes (rotation angle between parenthesis) applied to $^{1}$H and $^{13}$C (super indices) nuclear spins.

\begin{figure}[h!]
\includegraphics[width=0.48\textwidth]{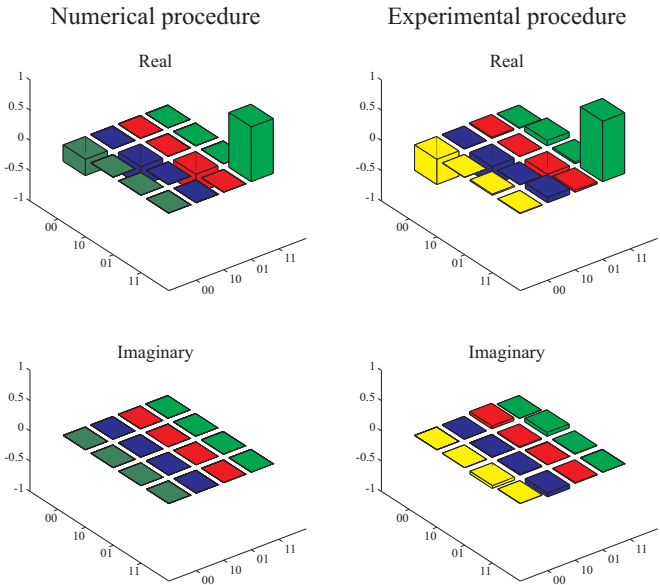}
\caption{Bar representation of the prepared deviation matrix. These states were produced by using hard pulses, free evolutions, and magnetic field gradients. The top (bottom) part of the figure corresponds to the real (imaginary) part of the deviation matrix while the left (right) part of the figure corresponds to the simulated (experimental) results for the prepared initial quantum state.}
\label{fig:EstadoInicialWitness}
\end{figure}

The pulse sequence (rf pulses, free evolutions, and magnetic field gradients) for preparation of the state $\left|11\right\rangle$ was numerically simulated and the real and imaginary parts of the deviation matrix are shown in the bar plots in the left part of Fig. \ref{fig:EstadoInicialWitness}. The experimental results, and corresponding pulse sequence, are displayed on the right part of Fig. \ref{fig:EstadoInicialWitness}. To obtain the experimental deviation matrix, the quantum state tomography procedure, which will be described in Sec. \ref{sec:levelC3}, was used. 

State preparation by spatial averaging can also be achieved replacing the hard rf pulses and free evolutions by numerically optimized low power pulses (known in the NMR terminology as soft pulses) followed by magnetic gradient pulses. Due to the excellent control of the NMR parameters that dictate the system evolution, the NMR Hamiltonian can be parametrized in terms of the rf pulses parameters (power $\omega_{1}$, duration $t$ and phase $\varphi$). This allow using numerical optimization routines to seek for the set of pulses, and the corresponding parameters, that brings the system from the thermal equilibrium to a state whose the diagonal element of the deviation matrix are the same as those of the target state. Thus, the subsequent application of the field gradient pulse vanishes all off-diagonal deviation matrix elements, producing the desired state. Two numerical procedures named as GRAPE (GRAdient Ascent Pulse Engineering) \cite{Khaneja2001,Khaneja2005}  and SMP (Strongly Modulated Pulses) \cite{fortunato2002} are commonly used for pulse optimization. An example of a SMP pulse sequence used to implement an specific initial quantum state is presented in Table \ref{tab:SMPvalues}. The first column indicates the order of application for a certain set of parameters. The second (fifth) and third (sixth) columns show the numerical values of the power and phase of the rf pulses applied to the $^{1}$H ($^{13}$C) nuclei. The duration of each rf pulse is shown in the fourth column. As already mentioned, after the RF pulse a magnetic field gradient is applied to wash out the off-diagonal coherences in the deviation matrix.  

\begin{table}
\begin{center}
\begin{tabular}{c||r|r|r|r|r}
\hline
SMP		& \multicolumn{5}{c}{ Parameters } \\ \cline {2-6}
Pulse	&   \multicolumn{2}{c|}{Hydrogen}  & time & \multicolumn {2}{c }{Carbon} \\ \cline{2-3} \cline{5-6}
Step	& $\omega_{1}^{H}$ & $\varphi^{H}$ & (ms) &    $\omega_{1}^{C} $    &    $\varphi^{C}  $     \\ \hline 		
	1		& 1092.6 & 5.94 &0.518 &1962.9 &1.88 \\
	2		& 1679.8 & 5.73 &1.546 &1276.7 &5.26 \\
	3		& 112.0  & 1.64 &0.226 & 619.3 &2.55 \\
	4		& 1998.0 & 2.63 &1.130 & 868.9 &3.91 \\
	5		& 211.3  & 4.84 &1.996 & 835.8 &5.46   \\ \hline
\end{tabular}				
\end{center}	
\caption{Numerical values of the power, duration and phase of the rf pulses utilized to implement the SMP technique for quantum state preparation. These parameters were used to produce diagonal initial quantum states (see, e.g., Fig. \ref{fig:PulseSequenceDynamics}). The SMP technique is one of the steps used to produce suitable quantum states for the study of quantum discord in NMR systems. For instance, using these values for the parameters characterizing the rf pulses we produced a Bell-diagonal state satisfying the following inequality: $\left|c_{1}\right|$, $\left|c_{2}\right|\geq\left|c_{3}\right|$. Similar procedures were followed to obtain other states of interest.}
\label{tab:SMPvalues}
\end{table}

\begin{figure}[h!]
\includegraphics[width=0.48\textwidth]{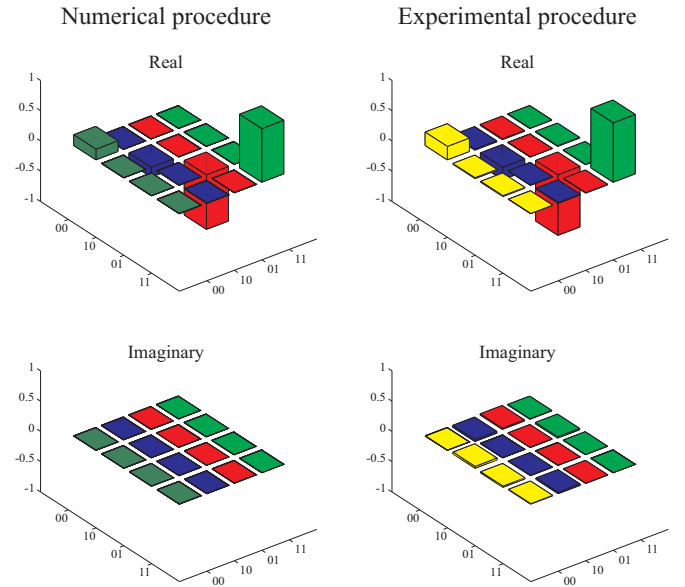}
\caption{Bar representation of the prepared deviation matrix. These states were produced by using the SMP technique. The top (bottom) part of the figure corresponds to the real (imaginary) part of the deviation matrix while the left (right) part of the figure corresponds to the simulated (experimental) results for the prepared initial quantum state.}
\label{fig:EstadoInicialDynamica}
\end{figure}

The application of the pulse sequence showed in Table \ref{tab:SMPvalues}, followed by the gradient pulse, to the thermal equilibrium state was numerically simulated being the corresponding real and imaginary parts of the deviation matrix shown in the left part of Fig. \ref{fig:EstadoInicialDynamica}. An experiment using the parameters values shown in Table \ref{tab:SMPvalues} was also performed. The resulting deviation matrix, also obtained using quantum state tomography, is shown in the right part of Fig. \ref{fig:EstadoInicialDynamica}.

The main advantage of using SMP or GRAPE pulses followed by magnetic field gradient pulses for state preparation via spatial averaging is their versatility in terms of producing optimized pulses that leads to any desired spin populations, which makes the method more general than the use of hard pulses, where a specific pulse sequence has to be found for each desired state.

The preparation of a quantum state via temporal averaging involves the realization of a specific number of time uncorrelated experiments. However, each experiment is designed in such a way that the resulting state obtained after summing up the result of all experiments, is the desired state. As in the case of spatial averaging, both the hard \cite{knill1998} and soft pulses \cite{fortunato2002,kampermann2005} approaches can be used. One advantage of this preparation scheme is the possibility of preparing non-diagonal states, that can be prepared directly from the thermal equilibrium, since no magnetic gradient pulses are applied. 

\begin{figure}[h!]
\includegraphics[width=0.48\textwidth]{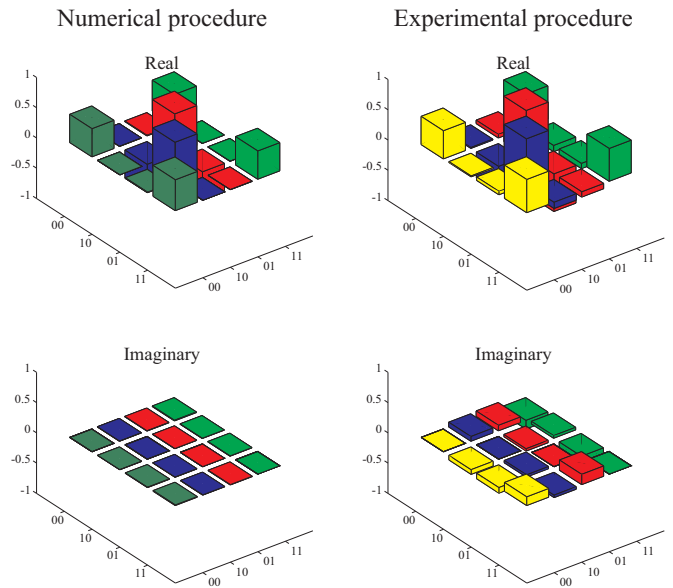}
\caption{Bar representation of the prepared deviation matrix. These states were produced by using hard pulses, free evolutions and magnetic field gradients. The top (bottom) part of the figure corresponds to the real (imaginary) part of the deviation matrix while the left (right) part of the figure corresponds to the simulated (experimental) results for the prepared initial quantum state.}
\label{fig:EstadoXQuadrupolar}
\end{figure}

An example of a state prepared using the temporal averaging scheme is shown in \ref{fig:EstadoXQuadrupolar}. This state was implemented using $^{23}$Na quadrupolar nuclei (spin $3/2$) in a lyotropic liquid crystal sample as described in the experimental section. A SMP pulse was also used in the state preparation, but in this case the parameter values were optimized based on the Hamiltonian of Eq. (\ref{HamiltonianoCristal}). Different experiments are labelled by the capital letters A, B, C, and D, each one having five sets of values for the rf pulses parameters ($\omega_{1}$, $t$ and $\varphi$), see Table \ref{tab:SMPvaluesSDS}. Each pulse was applied separately to the thermal equilibrium state, resulting in transformed deviation matrices, whose combination   (sum) is the final $\Delta\rho$. The final simulated $\Delta\rho$ is shown in the bar plot in the left of Fig. \ref{fig:EstadoXQuadrupolar}. The corresponding experimental $\Delta\rho$, obtained using same parameter values of Table \ref{tab:SMPvaluesSDS}, is shown in the right side of Fig. \ref{fig:EstadoXQuadrupolar}.

\begin{table}	
\begin{center}				
\begin{tabular}{c|c|r|r|r}
\hline
	Group	& SMP		& \multicolumn{3}{c}{ Parameters } \\ 
				& Pulse	&   \multicolumn{3}{c}{of Sodium nuclei}  \\ \cline{3-5}
	Set		& Step	& $\omega_{1} $ & $\varphi $ & time ($\mu$s)     \\ \hline 

		&	1 & 37485.4 & 6.208 & 15.11 \\
		&	2 & 11624.8 & 4.243 & 36.53 \\
	A	&	3 & 31867.5 & 3.293 & 4.15 \\
		&	4 & 38906.5 & 6.027 & 47.99 \\
		&	5 & 38681.0 & 1.927 & 31.62 \\  \hline 

		&	1 & 12063.7 & 3.919 &   7.95 \\
		&	2 & 15399.8 & 3.581 &  33.74 \\
	B	&	3 & 38721.2 & 5.845 &  30.47 \\
		&	4 & 24148.3 & 5.117 &   3.38 \\
		&	5 & 38916.9 & 3.762 &  48.08 \\  \hline 

		&	1 & 39208.2 & 1.788 &  38.82 \\
		&	2 &  7589.9 & 4.984 &  27.38 \\
	C	&	3 & 19817.7 & 0.889 &   9.27 \\
		&	4 & 38915.2 & 1.682 &  25.45 \\
		&	5 & 38755.6 & 1.200 &  21.98 \\  \hline 

		&	1 & 13372.2 & 4.333 &  15.24 \\
		&	2 &  2192.1 & 1.506 &  37.89 \\
	D	&	3 & 39094.7 & 4.433 &  29.35 \\
		&	4 & 11495.7 & 3.294 &  10.49 \\
		&	5 & 18867.5 & 0.213 &  36.92 \\ \hline

\end{tabular}
\end{center}
\caption{Numerical values of power, duration, and phase of the rf pulses used to prepare quantum states by implementing the SMP technique and the time average procedure. These pulse sequences were used to produce the diagonal initial quantum state (see Fig. \ref{fig:PulseSequenceDynamics}). The SMP technique is one of the steps needed to produce appropriate quantum states to study quantum discord in NMR systems. Using these parameter values we have produced quantum states that satisfy the following inequality: $\left|c_{1}\right|\geq\left|c_{2}\right|$, $\left|c_{3}\right|\neq 0$.}
\label{tab:SMPvaluesSDS}
\end{table}

\subsection{Quantum State Tomography\label{sec:levelC3}}

In general, the Quantum state tomography (QST) uses the observables of the quantum system to quantitatively measure each element of the density matrix that correspond to the state of the system. In NMR systems these observables are comprised by spin magnetizations, which depend on a combination of the density matrix that correspond to single quantum transitions ($\Delta m = \pm 1$). Thus, it is usually stated that only those elements located at certain reading position in the density matrix are experimentally accessed directly. Thus, in order to obtain the elements out of the reading positions it is necessary to apply a particular scheme of rotations to bring the desired elements to the reading positions in a controlled way. By doing so, the spin magnetizations become dependent on these "non directly detectable" elements, which can then be obtained after proper data processing. Thus, the art behind  QST is to find a minimum number of rotations that allow mapping the full sytem density matrix (actually the deviation matrix in the NMR context).  Indeed, there are many QST procedures in NMR QIP either using single or combined spin rotations\cite{long2001,leskowitz2004}, global rotations of the spin system \cite{teles2007}, transition selective excitations \cite{kampermann2005,bonk2004}, among others. Generally speaking, each  QST method is more appropriated for a specific spin system, for example, while the methods based on a set of single spin rotations are usually applied for spin $1/2$ systems, the methods based on transition selective pulses or global spin rotations are more appropriated for quadrupolar nuclei. 

\begin{table}[b]
\begin{center}
\begin{tabular}{c|c|c}  \hline
\textrm{Pulse order} & \textrm{Carbon} & \textrm{Hydrogen}\\
 \hline
1st & \textbf{I} & \textbf{I} \\
2nd & \textbf{X} & \textbf{X} \\
3rd & \textbf{I} & \textbf{X} \\
4th & \textbf{I} & \textbf{Y} \\
5th & \textbf{X} & \textbf{I} \\
6th & \textbf{Y} & \textbf{I} \\
7th & \textbf{X} & \textbf{Y} \\
8th & \textbf{Y} & \textbf{X} \\
9th & \textbf{Y} & \textbf{Y} \\  \hline
\end{tabular}
\end{center}
\caption{An ordered pulse sequence applied simultaneously in both nuclear species. $\textbf{X}\equiv \left(\frac{\pi}{2}\right)_{+x}$ represents a radio-frequency pulse of $\pi / 2$ in positive $x$-direction. $\textbf{Y} \equiv \left(\frac{\pi}{2}\right)_{+y}$ represents a radio-frequency pulse of $\pi / 2$ in positive $y$-direction. $\textbf{I}$ represents absence of radio-frequency pulse.}
\label{tab:tomographypulse}
\end{table}

\begin{figure}[h]
\includegraphics[width=0.48\textwidth]{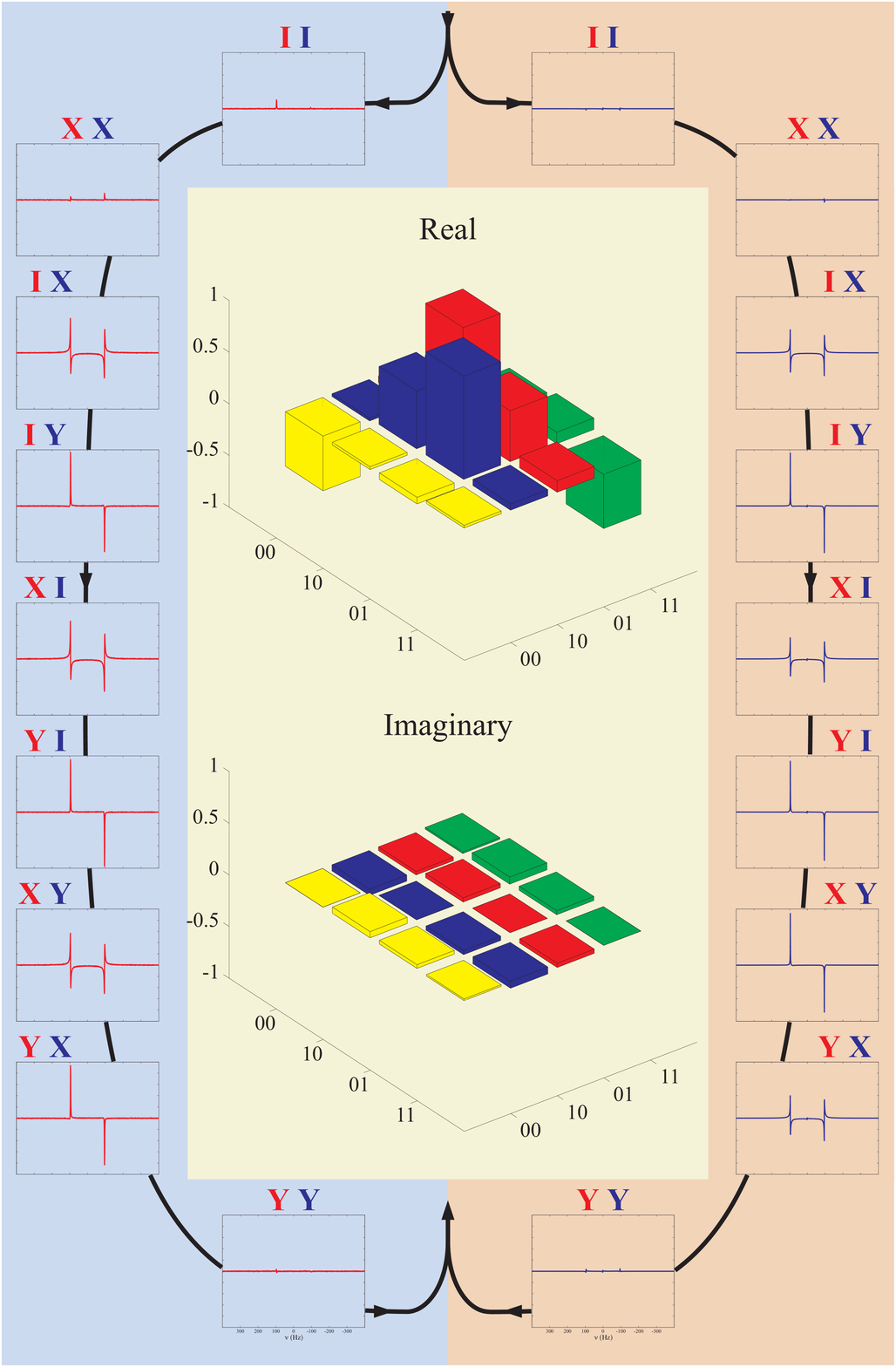}
\caption{Set of spectra of a two-spin system encoded in the chloroform molecule generated in the implementation of QST pulse sequence detailed in Table \ref{tab:tomographypulse}. Left (right) spectra corresponds to the Carbon (Hydrogen) nuclear spin. The bar plots in the middle corresponds to the real (top) and imaginary (bottom) parts of the experimentally deviation matrix obtained through quantum state tomography.}
\label{fig:TomografiaSequencia}
\end{figure}

As a specific example, we will briefly discuss here the QST technique introduced by Long \textit{et al.} in Ref. \cite{long2001}, a method widely used for QST in two-coupled spin $1/2$ system, which was also adapted for systems with higher number of spins \cite{leskowitz2004,cramer2010}. In this method, the QST is performed in 9 steps and, before each step, the system needs to be prepared in the same state, whose the deviation matrix need to be characterized. The first step consists in acquiring the NMR signal of both nuclei (free induction decay - FID) in the absence of radio-frequency pulses. After the Fourier transformation of the FID, the line intensities in the resulting spectrum, labelled as \textbf{II} in Fig. \ref{fig:TomografiaSequencia}, are measured and recorded. In each of the following steps, rf pulses with specific phases, as described in Table \ref{tab:tomographypulse}, are applied to each or both nuclei and the same reading procedure is executed. 

Fig. \ref{fig:TomografiaSequencia} shows in red (blue) the Carbon (Hydrogen) spectrum obtained after each tomography step, executed after preparing a Bell-diagonal state in the CHCl$_{3}$ two qubit spin system. The spectra are labelled as \textbf{II}, \textbf{XX}, \textbf{IX}, \textbf{IY}, \textbf{XI}, \textbf{YI}, \textbf{XY}, \textbf{YX}, \textbf{YY}, with \textbf{I} representing no pulse and \textbf{X} (\textbf{Y}) a $\pi / 2$ pulse applied in positive $x$ ($y$) direction. In this case, since each spectrum has two spectral lines with real and imaginary components, 72 line intensities are recorded. Each of this line intensities are associated with one or more elements of the deviation matrix, so to obtain the full deviation matrix one needs to know the relations between the line intensities (NMR read-outs) and the deviation matrix elements. Indeed, the rf pulse phases on \ref{tab:tomographypulse} are such that all the elements are accessible, with some redundancy for error minimization. 

To be more specific, let us consider the following deviation matrix,
\begin{equation}
\Delta \rho =\left[ 
\begin{array}{cccc}
a_{1} & a_{2}+ia_{11} & a_{3}+ia_{12} & a_{4}+ia_{13} \\ 
a_{2}-ia_{11} & a_{5} & a_{6}+ia_{14} & a_{7}+ia_{15} \\ 
a_{3}-ia_{12} & a_{6}-ia_{14} & a_{8} & a_{9}+ia_{16} \\ 
a_{4}-ia_{13} & a_{7}-ia_{15} & a_{9}-ia_{16} & a_{10}
\end{array}
\right],
\label{operadordensidade}
\end{equation}
which is described by sixteen unknown variables. The rotation imposed to the spin system by the rf pulses in each step $n$ of the tomographic process can be represented by the operator $\mathbf{U_{n}(X,Y)}$. Note that the analytical form of $\mathbf{U_{n}(X,Y)}$ is known for a specified pulse duration and phase. Hence, after the $n$-th rotation, the transformed deviation matrix is represented by  
\begin{equation}
\Delta \rho _{n}=\mathbf{U}_{n}\left( \mathbf{X,Y}\right) \Delta \rho \mathbf{U}^{\dagger }_{n}\left( \mathbf{X,Y}\right).
\label{TomografiaN}
\end{equation}

Thus, after the $nth$ transformation, the real ($M_{x}$) and imaginary ($M_{y}$) parts of the NMR magnetization for each nuclear spin {A,B}, become 
\begin{eqnarray}
M_{x,y;n}^{A} &=&\textrm{Tr} \left[\Delta \rho _{n} \left( \mathbf{I}_{x,y}\otimes\mathbb{I}\right)\right], \\
M_{x,y;n}^{B} &=&\textrm{Tr} \left[\Delta \rho _{n}\left( \mathbb{I}\otimes \mathbf{I}_{x,y}\right) \right], \label{eqmagn}
\end{eqnarray}

From Eq. \ref{eqmagn}, one can see that the magnetization values are linearly related with the $\Delta\rho$ elements, i.e., each $n$ value in Eq.\ref{eqmagn} represents a linear equation having the $a_{j}$'s $\Delta\rho$ elements as independent variables, the magnetizations $M_{x,y;n}^{A,B}$ as dependent variables and the elements of the operator $\mathbf{U_{n}(X,Y)}$ as coefficients. Note that the magnetization values are experimentally accessed and directly related to the line intensities and the $\mathbf{U_{n}(X,Y)}$ elements are known from the rf pulse parameters. Thus, Eqs. \ref{eqmagn} can be rewritten as 
\begin{eqnarray}
{b}_{k} &=&{X}_{kj} {a}_{j}, 
\label{ecuaciones73}
\end{eqnarray}
where $b_{k}$ with $k=1,...,73$ state for the $k$-th recorded line intensity refer to the $\Delta\rho$ elements before the transformation and $j=1,...,16$. The elements $X_{kj}$ are obtained from the pulse parameters applied in each QST step. 

The solution of equations (\ref{ecuaciones73}) for the ${a}_{j}$ variables enable us to reconstruct the deviation matrix. An example of this reconstruction procedure is presented in the bar-graph of Fig. \ref{fig:TomografiaSequencia}.

The QST methods for quadrupolar nuclei follows an analogous idea, i.e., obtaining a series of read-outs after applying specific rf pulses, recording the corresponding line intensities and solving a linear equation system to obtain the deviation matrix elements. However, the specificities of the method are somewhat more intricate and will not be presented here. More information can be found in Ref. \cite{teles2007}.

\subsection{Experimental Measurments of Quantum Correlations in NMR systems\label{sec:levelC5}}

In this section, experimental investigations regarding the quantification and identification of quantum correlations in NMR systems at room temperature will be presented. At first, we verify that the peculiar dynamics of quantum discord under decoherence, theoretically predicted for phase-noise channels \cite{Maziero-SC}, can be present even under the influence of an additional thermal environment \cite{Auccaise-SC}. Next, we present the first experimental implementation of a classicality witness \cite{Maziero-W}, which was used to identify the nature of the correlations in NMR systems without the necessity of a full QST \cite{Auccaise-W}.

\subsubsection{Experimental Dynamics of Correlations\label{sec:levelC6}}

Here we present an experimental verification of the sudden change phenomenon of quantum discord under decoherence, whose theory was discussed in Section \ref{SC-theory}. Although, in principle, such kind of phenomenon may take place in the quadrupolar system in a liquid-crystal sample, introduced Section \ref{sec:levelC1B} \cite{Diogo-Quadrupolar}, it is very difficult to observe it experimentally. The reason is that in this system the phase noise environment is global \cite{Diogo-Redfield}, and the transversal relaxation time is very short, particularly when compared with that for the two nuclear spins-1/2 in a liquid sample  (see Section \ref{sec:levelC1A}). Thus, for experimentally demonstrating the peculiar behaviour of correlations under decoherence, a two-qubit system encoded in the $^{1}\mathrm{H}$ and $^{13}\mathrm{C}$ nuclear spins in a $CHCl_{3}$ molecules, as discussed in section \ref{sec:levelC1A}, was used. As already mentioned, in this system the relaxation process that causes phase decoherence and energy dissipation is due, mainly, to internal molecular or atomic motions that leads to random fluctuations in the electromagnetic field in which the qubits are immersed. The decoherence process in this system can be modelled using the composition of a phase-damping channel with a generalized amplitude damping channel:
\begin{equation*}
 \rho_{AB}(t)=\mathcal{E}_{p}\circ\mathcal{E}_{a}(\rho_{AB}),
\end{equation*}
where the quantum operations $\mathcal{E}_{n}$ ($n=p,a$) are written in the operator-sum representation as shown in Section 1, i.e.
\begin{equation*}
 \mathcal{E}_{n}(\rho)=\sum_{j}K_{j}^{n}(t)\rho \left[K_{j}^{n}(t)\right]^{\dagger}.
\end{equation*}
where $K_{j}^{n}(t)$ are the well-known Kraus' operators. For the \textit{generalized amplitude-damping channel} the Kraus' operators are given by
\begin{subequations}
\begin{align*}
K_{0}^{a}  &  =\sqrt{\gamma}
\begin{pmatrix}
1 & 0\\
0 & \sqrt{1-p}
\end{pmatrix}
,\quad K_{1}^{a}=\sqrt{\gamma}
\begin{pmatrix}
0 & \sqrt{p}\\
0 & 0
\end{pmatrix}
,\\
K_{2}^{a}  &  =\sqrt{1-\gamma}
\begin{pmatrix}
\sqrt{1-p} & 0\\
0 & 1
\end{pmatrix}
,\quad K_{3}^{a}=\sqrt{1-\gamma}
\begin{pmatrix}
0 & 0\\
\sqrt{p} & 0
\end{pmatrix},
\end{align*}
\end{subequations}
where, in the NMR context, 
\begin{equation}
\gamma \approx \frac{1-\varepsilon}{2} 
\end{equation}
and 
\begin{equation}
 p=1-\exp\left(-t/T_{1}\right),
\end{equation} 
with $T_{1}$ being the longitudinal relaxation time of the qubit under consideration. In our case, each qubit have distinct Larmor frequencies, leading to also distinct relaxation times. The measured spin-lattice relaxation times are
\begin{equation*}
 T_{1}(^{1}\textrm{H})=2.5\mbox{ s} \text{ and } T_{1}(^{13}\textrm{C})=7\mbox{ s}.
\end{equation*}
The Kraus' operators for the \textit{phase-damping channel} are given by Eqs. (\ref{phase}) with $p=1-\exp\left(-t/T_{2}\right)$, where the measured transverse relaxation times associated with the the two qubits are given by
\begin{equation*}
T_{2}(^{1}\textrm{H})=1.8\mbox{ s} \text{ and } T_{2}(^{13}\textrm{C})=0.29\mbox{ s}. 
\end{equation*}
As no refocusing pulse was used, the effective transverse relaxation times are
\begin{equation*}
 T_{2}^{*}(^{1}\textrm{H})=0.31\mbox{ s} \text{ and } T_{2}^{*}(^{13}\textrm{C})=0.12\mbox{ s}.
\end{equation*}

\begin{figure}[h!]
\includegraphics[width=0.48\textwidth]{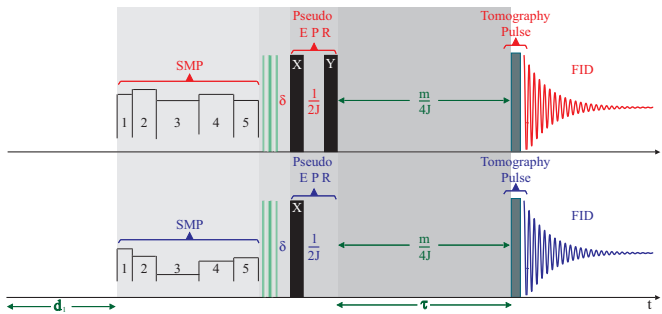}
\caption{Description of pulse sequence used to implement the dynamics of quantum and classical correlations under decoherence. The bottom (top) sequence corresponds to the representation of pulses applied to the $^{1}$H ($^{13}$C) nuclei. At first, the SMP technique and a $z$-gradient pulse are applied to implement a diagonal state. In the sequence, a pseudo-EPR gate is performed. In the third step, the nuclear system and environment interact resulting in decoherence of quantum states. Finally, the read-out of the quantum state is performed for time intervals of $m/4J$ ($m=0,1,2,...,250$) using the quantum state tomography procedure.}
\label{fig:PulseSequenceDynamics}
\end{figure}

\begin{figure}[h!]
\includegraphics[angle=0,width=0.50\textwidth,height=0.25\textheight]{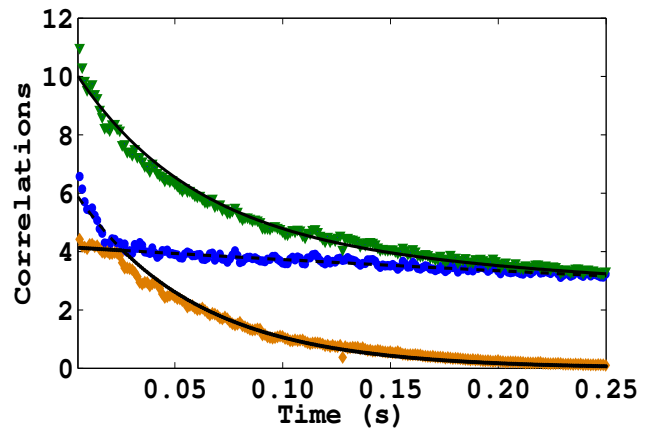}
\caption{Sudden change in behaviour and robustness of classical and quantum correlations under decoherence. The green downward triangles are the experimental data for the quantum mutual information, while the blue circles and orange upward triangles represent the classical and quantum correlations, respectively. The black lines are the theoretical predictions. The initial state is analogous to the state in Eq. (\ref{eq:Bell-D}) with $|c_{1}|,|c_{2}|>|c_{3}|$. The correlations are displayed in units of $(\varepsilon^{2}/\ln2)\mathrm{bit}$.}
\label{fig:correlations-dynamics}
\end{figure}

To verify the sudden-change in the decay rate and the robustness of correlations to phase-noise environments, a suitable initial state was prepared by mapping the NMR deviation matrix $\Delta\rho$ into a Bell-diagonal  state (Eq. (\ref{eq:Bell-D})) with $|c_{1}|,|c_{2}|>|c_{3}|$. An illustration of the pulse sequence utilized for accompanying the correlations dynamics is shown in Fig. \ref{fig:PulseSequenceDynamics}. Using the deviation matrices obtained from quantum state tomography, the correlations were numerically computed for each time step $m/4J$. The results are presented in Fig. \ref{fig:correlations-dynamics}, where the sudden-change in the decay rate of correlations is clearly present even in the presence of the strong thermal noise produced by the fluctuating fields of the nuclear spin environments. As already discussed the classical correlation should be constant under a phase damping, but a small decay is observed in Fig. \ref{fig:correlations-dynamics},which is due to the presence of the thermal environment. We see that the theoretical predictions presented are in good agreement with the experiment. We also verify the transition between two decoherence regimes \cite{Mazzola}. In the first regime the decoherence affects more strongly the classical aspect of correlation, and the quantum correlations have a small decay rate. After the sudden-change point, the classical correlation becomes more robust against decoherence and the noise affects more the quantum discord. It is worthwhile to note that the experimental deviation matrices have small coherences, which oscillate during the evolution. This leads to the small oscillations in the spectral line intensities used to perform the state tomography and, because the tomography process involves solving coupled equations, this oscillatory behaviour causes the oscillations observed in the correlations.

\subsubsection{Experimental Implementation of a Classicality Witness\label{sec:levelC7}}

\begin{figure}[h!]
\includegraphics[width=0.47\textwidth]{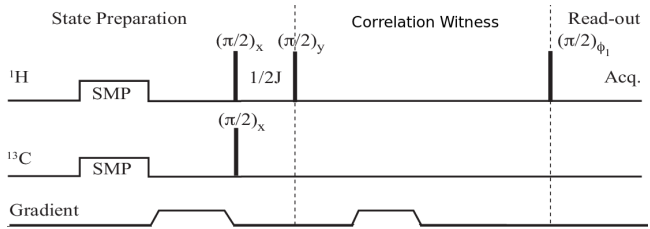}
\caption{Description of pulse sequence used to implement the Classicality Witness. The bottom (top) sequence corresponds to the representation of the pulses applied to the $^{1}$H ($^{13}$C) nucleus. At first, the SMP technique and a $z$-gradient pulse are used to implement a diagonal state. In the second step a pseudo-EPR gate is implemented. The third step is devoted to the interaction of the system with the environment to produce de decoherence of quantum states. Finally the read-out of the quantum state was performed at $m/4J$ time intervals with $m=0,1,2,...,11$, using the tomography procedure.}
\label{fig:PulseSequenceWitness}
\end{figure}

As we discussed so far, the quantification of quantum correlations involves quantum state tomography. However, in some situations it is enough to identify the nature of correlations in a system. To meet this goal one usually uses correlations witnesses. It turns out, as noticed in Section 2, that the space of classical states is not convex. For that reason, linear witnesses cannot be used to identify the quantumness of correlations in separable states. In what follows we present an experimental implementation of the nonlinear classicality witness introduced in Section \ref{sec:class-wit} \cite{Maziero-W}. The experiment was performed using the same NMR apparatus utilized to verify the sudden-change phenomenon of quantum discord under decoherence \cite{Auccaise-W}. The pulse sequence used to implement the protocol for the classicality witness is sketched in Fig. \ref{fig:PulseSequenceWitness}. Three states were prepared, a thermal equilibrium state $\rho_{T}$, a quantum correlated state $\rho_{QC}$ and a classically correlated stated $\rho_{CC}$. The measured values of the witness for three different states is presented in Table \ref{witness-table}. The thermal state has $W_{\rho_{T}}=0.05$, so this is assumed as the classicality cutoff limit of our experiments . For the classical state $W_{\rho_{CC}}=0.04$, which is within the classicality cutoff limit. The quantum-correlated state has $W_{\rho_{QC}}=3.13$, far above the classicality bound $0.05$. As a matter of fact, the classicality witness perfectly sorts out quantum and classically correlated states. Table \ref{witness-table} also shows the quantum discord computed from the experimentally reconstructed deviation matrices, which is in agreement with the classicality witness. However quantum discord is obtained after quantum state tomography followed by numerical extremization procedures.  

\begin{table}[h!]
\caption{Witness, quantum discord, and classical correlation measured in three different initial states: quantum correlated $\rho_{QC}$, classically correlated $\rho_{CC}$ and thermal equilibrium $\rho_{T}$. The witness was measured directly by performing the sequence of pulses depicted in Fig. \ref{fig:PulseSequenceWitness}, while the classical correlation and the symmetric quantum discord was computed after full QST and numerical extremization procedures. The correlations are displayed in units of $(\varepsilon^{2}/\ln2)\textrm{bit}$.}
\label{witness-table}
\begin{center}
\begin{tabular}{l|ccc}
                       &  $\rho_{QC}$   &  $\rho_{CC}$  &  $\rho_{T}$      \\ \hline
Witness                &    3.13        &   0.04        &  0.05            \\ 
Quantum Discord        &    4.02        &   0.00        &  0.00             \\ 
Classical Correlation  &    2.09        &   7.15        &  0.00             \\ 
\end{tabular} 
\end{center}
\end{table}

\begin{figure}[h!]
\includegraphics[angle=270,width=0.48\textwidth]{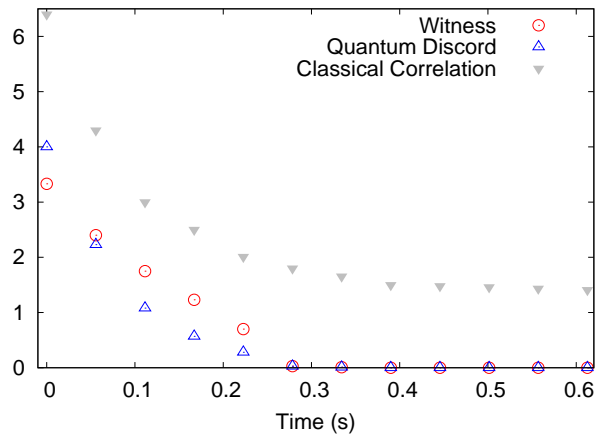}
\caption{Measured witness and computed correlations for $\rho_{QC}$ relaxed during a time interval $t_{n}=n\delta t$ ($\delta t=55.7\mathrm{ms}$, $n=0,1,2,\cdots,11$), before performing the witness measurement protocol. The red circles represent the witness, the downward gray triangles are the amount of classical correlation, and the blue upward triangles represents the symmetric quantum discord. The correlations are displayed in units of $(\varepsilon^{2}/\ln2)\textrm{bit}$.}
\label{fig:witness-dynamics}
\end{figure}

At last, we investigated the decoherent dynamics of the witness. We let the system, prepared in the initial state $\rho_{QC}$, evolve freely during a time period $t_{n}=n\delta t$ ($\delta t=55.7\mathrm{ms}$, $n=0,1,2,\cdots,11$) under the action of the NMR environment. The experimental results are shown in Fig. \ref{fig:witness-dynamics}. In this figure we also show the values of the correlations. We obtain a fairly good agreement between the witness values and the correlation quantifiers when identifying classical states.

\section{Concluding Remarks}

Quantum information processing has the potential of providing great advances, not only on the foundations of physics, specially to quantum mechanics and thermodynamics, but also on the technological ground. Therefore, studying the basis of quantum information is of fundamental importance.

Quantum correlations is thought to be responsible for the gain offered by quantum mechanics when compared with classical protocols. Our understanding about what is quantum and what is classical in a quantum correlated system have changed dramatically since the seminal EPR paper in 1935 \cite{EPR}. Initially, we thought that quantum correlations were the same as quantum non-locality. After that, entanglement was raised to the status of quantum correlations. In the last decade we realized that there are quantum correlations present even in separable (non-entangled) states and that such correlations could be used to over-perform classical systems in quantum information protocols. The most popular quantifier for such kind of correlation is called quantum discord \cite{Ollivier-Zurek}.

In this review we discussed the theoretical and the experimental aspects of quantum discord in nuclear magnetic resonance systems at room temperature. After discussing the dynamics of quantum discord under decoherence, we presented a witness for such a correlation. The experimental investigations of both issues was then treated, with a brief review of control techniques in NMR as well as of quantum state interrogation, a procedure known as quantum state tomography.

\begin{acknowledgements}
The authors acknowledge financial support from UFABC, CNPq, CAPES, FAPESP, FAPERJ and Brazilian National Institute of Science and Technology for Quantum Information (INCT-IQ). The warmly hospitality of Instituto de F\'{i}sica de S\~ao Carlos - Universidade de S\~{a}o Paulo (IFSC-USP) and Centro Brasileiro de Pesquisas F\'{i}sicas (CBPF), where the experiments were performed, is also acknowledge.
\end{acknowledgements}

\end{document}